# Power and limitations of electrophoretic separations in proteomics strategies

Thierry. Rabilloud [1,2], Ali R.Vaezzadeh [3], Noelle Potier [4], Cécile Lelong [1,5], Emmanuelle Leize-Wagner [4], Mireille Chevallet [1,2]

1: CEA, IRTSV, LBBSI, 38054 GRENOBLE, France.
2: CNRS, UMR 5092, Biochimie et Biophysique des Systèmes Intégrés, Grenoble France
3: Biomedical Proteomics Research Group, Central Clinical Chemistry Laboratory, Geneva University Hospitals, Geneva, Switzerland
4: CNRS, UMR 7177. Institut de Chime de Strasbourg, Strasbourg, France
5: Université Joseph Fourier, Grenoble France

Correspondence :

Thierry Rabilloud, iRTSV/LBBSI, UMR CNRS 5092,

CEA-Grenoble, 17 rue des martyrs,

F-38054 GRENOBLE CEDEX 9

Tel (33)-4-38-78-32-12

Fax (33)-4-38-78-44-99

e-mail: Thierry.Rabilloud@ cea.fr

Abstract:
Proteomics can be defined as the large-scale analysis of proteins. Due to the complexity of biological systems, it is required to concatenate various separation techniques prior to mass spectrometry. These techniques, dealing with proteins or peptides, can rely on chromatography or electrophoresis. In this review, the electrophoretic techniques are under scrutiny. Their principles are recalled, and their applications for peptide and protein separations are presented and critically discussed. In addition, the features that are specific to gel electrophoresis and that interplay with mass spectrometry( i.e., protein detection after electrophoresis, and the process leading from a gel piece to a solution of peptides) are also discussed.

Table of contents

# I. Introduction

Among the variety of available spectroscopic techniques, the combination of speed, sensitivity, and accuracy makes mass spectrometry an obvious choice in proteomics strategies; i.e., strategies that aim at the wide-scale characterization of proteins and protein variants in biological samples.

However, mass spectrometry alone cannot solve this analytical problem of the wide-scale characterization of proteins, for several major reasons. First of all, biological samples often contain several hundreds to several tens of thousands of different protein forms. The **word of "protein forms"** encompass, of course, gene products with all of their post-translational modifications (PTM); e.g., glycosylation, phosphorylation, or protein cleavage. On top of the diversity problem, the dynamic range of protein presence ( i.e., the mass ratio between the rarest protein form and the most abundant ones) is often far beyond the quantitative dynamic range of a mass spectrometer. For example, this dynamic range covers 4 orders of magnitude in prokaryotic samples, 6 in eukaryotic cells (Lu et al., 2007), and 12 in complex biological fluids such as plasma (Anderson and Anderson, 2002).

Still going on this trend of complexity, the analyte itself (i.e, the protein) is very complex, so that even the most precise mass measurement of a complete, modified protein often does not give an unequivocal answer, in the sense that several modified proteins can fit within the same mass measurement window.

Last but not least, it is quite frequent in biology to have to address quantitation problems. A good example is provided by blood tests, in which normal and pathological values are defined, with the normal values frequently not zero. Thus, a quantitative approach is often needed, and this goal is not always easy to implement for mass spectrometry. However, the discussion of the quantitation issues in mass spectrometry for proteomics is clearly outside the scope of this review.

Because of the complexity of the analytes, the mass spectrometric measurement is not carried out on intact proteins, but on smaller peptides that are produced from the proteins by a controlled proteolytic process. The multiple measurements made on peptide masses, either on a few peptides by MS/MS or on a combination of peptides that arise from the same protein (e.g., for peptide mass fingerprinting), allow an unequivocal characterization of the proteins of interest and sometimes of some post-translational modifications.

Because of the complexity of biological samples, it is absolutely necessary to separate (and sometimes to quantify) the analytes prior to their measurement with mass spectrometry, so that the complexity of what is introduced into the mass spectrometer is compatible with the performance of the instrument in the chosen measurement mode. This separation can be made on the proteins of the sample or on the peptides that arise from protein digestion (or on both). Among the various separation modes available for proteins and peptides, chromatography and electrophoresis are almost exclusively used at the current time. Thus the purpose of this review is to provide to the reader a critical review of the input of electrophoretic techniques in a proteomics strategy.

# II. The principles at play

By definition, electrophoresis consists of the separation of analytes (here peptides or proteins) as ions that are driven differently (in order to achieve separation) under an electric field. Two modes of electrophoresis are mostly used for peptide and protein separation; zone electrophoresis and isoelectric focusing.

In zone electrophoresis, a speed race is made in the separation medium, usually at constant pH, and the speed of each analyte is dictated by its charge, which drives it under the electric field, and by the friction forces, which slow it down. In order to minimize the impact of diffusion, which always tends to broaden the peaks of analytes, and also to minimize the impact of the volume of the initial sample on the final resolution, a special trick, called discontinuous electrophoresis, is used (Davis, 1964), schematized on figure 1. In this system, a composite separation medium with two phases is used, (figure 1, left panel). The sample is loaded on top of the first phase, called the concentration phase, where an isotachophoresis is carried out. Isotachophoresis (same speed electrophoresis) uses the rules of transport of ions in an electric field to build what is called a moving boundary, in which the ions are ranked in the order of their mobility. Under proper conditions, all analyte ions can be included in this moving boundary, whose dimensions are dictated by the ionic strength of the medium and by the concentration of each ion (for detailed calculations, see Jovin (Jovin, 1973a, Jovin, 1973b). By sweeping this moving boundary through the sample and then through the concentration phase of the separation medium, all analytes can be concentrated in the moving boundary (figure 1, center panel). Although this produces a spectacular increase in resolution, the concentration zone reflects its name very adequately, because the analytes are extremely concentrated in the moving boundary, with the risk that they exceed their solubility and start to precipitate. But, by definition, no macroscopic separation of the analytes takes place in the concentration phase. Separation takes place in another phase, called the separation phase, in which the speed of the moving boundary is increased to exceed the mobility of the analytes, which are left behind and separated by zone electrophoresis in the trailing phase (figure1, right panel).

Isoelectric focusing exploits special properties of some complex ions, which bear both weak acid and weak bases, which is the case of peptides and proteins via the amino and carboxy termini and by the side chains of some amino acids. Such complex ions have an isoelectric point, also noted pI (i.e. a pH at which the positive and negative charges are equal) so that the resulting electric charge of the analyte is zero. Below the isoelectric point (i.e. at a more-acidic pH) the analyte will be a cation, and above the isoelectric point (i.e. more-basic pH) it will be an anion.

The basis of isoelectric focusing is to use a separation medium in which a smooth pH gradient that encompasses all of the pIs of the analytes of interest (figure 2A). If the acidic end of the gradient is connected to the + electrode and the basic end to the - electrode, then the analytes will be separated according to their pIs, in the sense that each analyte, wherever it is applied in the pH gradient, will migrate to its steady-state position represented by its pI (figure 2B). The method is called isoelectric focusing because each analyte is concentrated at its pI by the counteracting forces induced by diffusion (which tends to broaden the analyte zone) and by the electric field (which tend to drive back the analyte at its exact pI) (see figure 2C). More detailed explanations on isoelectric focusing can be found in dedicated books (Righetti, 1983).

Opposite to the situation in chromatography, the separation in electrophoresis is not strictly dependent on the presence of a solid phase. However, the choice of electrophoresis configurations is indirectly limited by the Joule effect. The heat induced by an electric current provokes important convection flows, which limit in turn the resolution by blurring the separations produced upon electrophoresis. Thus, adequate resolutions are obtained only when convection flows are limited, and this limitation is achieved mostly by two very different setups.

The most obvious setup, but not the most extensively-used one, is to use a capillary fluid column for separation. Because of the minimal cross-section, minimal electric power is used, and thus minimal heating is produced.

Although such a setup could be, in principle, a real alternative to chromatographic columns for the separation of proteins or peptides, it has never gained widespread use, although there are some publications in the field. This limited use is due to a variety of reasons, among which two prominent ones are worth mentioning.
First, the capillary electrophoresis separation of complete proteins has always been plagued by adsorption on the walls, which has prevented any robustness in separation
Second and more importantly, the loading achievable in capillary electrophoresis is too low to be of practical interest in proteomics studies. Indeed, proteomics studies imply that one can analyze at the same time rather abundant analytes and rather rare ones in the same sample. That requirement implies in turn that one can load rather high amounts of sample. Such loadings are achievable in column chromatography by recirculating the sample in a loading cartridge ,which is pulsed into the capillary column, but this preconcentration process is not so easy to achieve with a capillary electrophoresis setup, despite the use of stacking processes (Locke and Figeys, 2000), solid-phase extraction (Figeys et al., 1998) or isoelectric focusing and its built-in concentrating effect (Simpson and Smith, 2005, Yu et al., 2006, Wang et al., 2007).

All these reasons explain why the second setup for limiting convection ( i.e. gel electrophoresis) has gained the most widespread use. As its name says, it consists of conducting the electrophoretic separation in a conductive gel; i.e,. a mixed buffer-crosslinked polymer system that has pores of a more or less defined size. The polymer chains that delimit the pores break the convection flows, and ensure adequate resolution. Moreover, they also limit the diffusion, which thereby further enhances resolution. Last but not least, the friction of the analytes on the gel material adds another separation parameter that can be used to enhance the resolution of the complete technique, especially when the size of the analyte is the separating parameter.

In addition to these advantages in separation, working in a gel medium has other important advantages related to the use of solid phase chemistry, but these features are detailed later in this paper (section IV.B).

However, the price to pay for the use of gel media is the loss of automation, but this loss is compensated, at least in part by the fact that gel separations are continuous separations and can be parallel ones, either handling multiple samples on one gel, or providing a continuous two-dimensional process. This latter aspect is worth discussing in more detail:

When a liquid separation is carried out, and when the separated analytes must be further analyzed in another liquid (or gel) separation process, fractions must be collected. When a few analytes are separated, this fraction collection process can be data-dependent, thereby isolating each analyte peak in a limited number of fractions. However, when very complex samples are separated, as is the case in proteomics, data-dependent fraction collection becomes pointless, because it is obvious that any fraction collection scheme will split at least some peaks in two. This peak splitting means in turn that the fraction-collection process that is mandatory in liquid separations introduces artefactual discontinuities in the separations, and splits some analytes into multiple fractions. A good visual illustration of this phenomenon can be found in schemes that couple liquid chromatography to gel electrophoresis (Szponarski et al., 2004, Szponarski et al., 2007, Schluesener et al., 2007).

Conversely, two-dimensional gels, whatever the separation principles used in both dimensions, represent a happy exception to this rule, because the first-dimension gel, which represents a continuous separation of the analytes, can be loaded as a whole on the top of second dimension gel, to provide a completely continuous two-dimensional separation, therefore without artefactual discontinuity. This continuity in separation, results in higher resolution and higher reproducibility of the two-dimensional gel-based separations.

Of course, there are constraints on the gels that can be used in electrophoretic separations. Such gels should have pores of an adequate and tunable size, to allow either non-restrictive gels or restrictive gels to be used. More importantly, such gels should be devoid of electric charges, which would otherwise cause strong electroendosmosis that complicates or disturb the separation. They should also not adsorb proteins, because adsorption would cause tailing of the separations. In the frame of proteomics studies, these gels should also resist the organic solvents used in the washing, digestion and extraction steps.

This series of constraints have confined the gels used in electrophoresis to crosslinked polyacrylamide. Because of its crosslinked nature, this polymer is stable to many adverse conditions. However, it is not very stable in strongly basic conditions, which lead to the hydrolysis of the amido group. The gel characteristics are also altered by organic solvents (Righetti 1989), which limits the scope of use of this gel. For these reasons, alternate acrylic monomers have been proposed (Artoni et al., 1984, Zewert and Harrington, 1992, Chiari et al., 1994, Simò-Alfonso et al., 1996), but none of them has gained any practical widespread use.

It should be also mentioned that polyacrylamide gels are polymerized in situ, and that process poses additional problems. As a matter of facts, only a fraction of the monomers is polymerized when gelification occurs (Righetti and Caglio, 1993); however the monomer conversion in the polymer continues after gelification to yield much better final incorporation yields (Gelfi and Righetti, 1981a, Gelfi and Righetti, 1981b). However, even with a 90% final conversion rate, the free acrylamide monomer exists in concentrations in the 0.1M range, and this free monomer can react with proteins side chains (Chiari et al., 1992), to lead to complications in the proteomics experiments.

Moreover, acrylamide-based gels can be used to devise immobilized pH gradients (Rosengren et al., 1977) in which the gradient is formed by copolymerization of buffering acrylamide monomers with the standard acrylamide and crosslinker. This way of creating the pH gradient mandatory for isoelectric focusing has revolutioned the scope and use of IEF. Detailed description of the many advantages of immobilized pH gadients is far beyond the scope of this paper, and can be found in monographies (Righetti, 1990).

Although gel-based separations represent by far the most important part of electrophoretic separations used in proteomics studies, some non-capillary, gel-free systems are sometimes used, most often in a complex separation scheme where they are used in an upstream part, with gel electrophoresis being often the downstream part. When confining the point to proteins and peptide separations, and not to pre-separations of biological samples (e.g. , Zischka et al., 2003), these schemes use almost always isoelectric focusing (Hoffmann et al., 2001, Xie et al., 2005) because this type of separation is less sensitive to convection for several reasons. First of all, IEF operates at very low ionic strength, thereby limiting the Joule effect. Second, IEF is a steady-state process, in which molecules moved away from their steady-state position will be driven back to it by the electric field. Thus, although several devices for free-flow isoelectric focusing of molecules have been described in the literature, those that have been used in proteomics studies belong mainly to the segmented systems, either using ampholytes to build the pH gradients (Egen et al. 1988), (Shang et al., 2003), or using isolectric membranes (Righetti et al., 1987, Righetti et al., 2005, Han et al., 2008), or a derivative thereof (Ros et al., 2002). Another variant of these IEF-based pre-fractionation techniques is represented by the Sephadex IEF (Radol,a 1973, Radola, 1975, Görg et al., 2002), in which the IEF process is conducted in a layer of granulated beads.

## III. How to use electrophoresis in a proteomics strategy

As stated in the introduction, the analytes in a proteomics strategy are by definition proteins, or the peptide that arise from their natural or provoked degradation. Consequently, the application of electrophoresis in proteomics can be split into two subfields, which are, respectively, the separation of proteins and the separation of peptides.
 However, a central point crosses both subfields, and is found in all "omics". In contrast to what is done in classical biochemistry, where the separation processes are optimized for each and every protein under study, the large-scale approaches rely on a "one size fits all" dogma. Even if this universality is not possible, at least in the proteomics world, the approach is still a "one size fits many" approach. In other words, the separation techniques that can be successfully used in proteomics must be very generic and wide-scope.
 Furthermore, proteomics analysis does not stop at this separation stage. Thus, the separation techniques of interest must be compatible with the downstream analysis, which is now almost always carried out by mass spectrometry. Thus, the resulting sample must be either free of substances that interfere with mass spectrometry (e.g. salts, polymeric reagents, surfactants) or at least cleanable by a method that is compatible with mass spectrometry, generally reverse phase chromatography.
 These two constraints, genericity and compatibility with downstream analysis, limit the types of separations that can be used, and electrophoresis is no exception to this rule.

## III.A. Electrophoretic separation of peptides.

Although peptides are a much simpler analyte than proteins, their separation by electrophoresis in proteomics has been described much later than the one of proteins. This delayed used of electrophoresis is due to the fact that the classical electrophoretic techniques, and especially zone electrophoresis, are not well- suited to peptide separation. As a matter of fact, adaptation of zone electrophoresis to peptide separation requires a number of adjustments (West et al., 1984, Schägger and Von Jagow, 1987), and even though the resolution is not sufficient to be useful in the 500-3000 Da range, where most of the peptides arising from a complete protein digestion will be. Although the size of the peptides is not a very good separation parameter, parameters that depend importantly on amino-acid composition are much more useful. Consequently, within the electrophorestic separations, isoelectric focusing is the separation of choice for peptides. However, the generation of the pH gradients needed in IEF has for long relied on the use of carrier ampholytes. These ampholytes molecules are also polycarboxylic polyamino compounds, and thus structurally related to peptides in terms of size and of chemical functions. This structural relation makes the separation between the bulk of carrier ampholytes and the small amounts of peptides difficult, and this contamination of peptides by ampholytes hampers further peptide analysis. This ampolytes-associated problem explains why the separation of peptides in proteomics experiments are most often carried out by immobilized pH gradients, which offer of much cleaner separation in terms of chemical noise.
This example is a typical illustration of the constraints imposed by the downstream analysis. Because the peptides are the real analyte to be studied by mass spectrometry, the separation system must avoid to pollute them too heavily. Otherwise, sample cleaning would be difficult and/or complicated, with associated losses of some peptides and thus of information.

The only exception to this rule is capillary IEF, in which the pH gradient is always generated by ampholytes. However, because of the very low volumes in capillary IEF, the mass contamination by ampholytes is reduced to a more acceptable level.

### III.A.1. Capillary IEF

Capillary IEF (CIEF) is a high-resolution separation technique that can be applied for amphoteric compounds, such as proteins or peptides. These analytes are separated according to their pI values, in a pH gradient formed under the influence of an electric field. CIEF was the first IEF technology used for analysis of peptides. Already, in the 1990s there were several groups that used CIEF for peptide fractionation and identification via UV detection or by MS. Mazzeo provided the first report of peptide mapping by CIEF (Mazzeo *et al,*. 1993). Bovine and chicken cytochrome c were fractionated, and the theoretical and calculated pIs were compared. They found an acceptable correlation between the two values. The major limitation of their method was that, because UV detection at 280 nm was employed, only tryptophan- and tyrosine-containing peptides were detected. Detection at 280nm was selected because the ampholytes used to generate the pH gradient absorb below 280 nm. A year later, Vonguyen *et al.* published a similar research (Vonguyen *et al,*. 1994). Kasai's group demonstrated the utility of synthetic peptides as isoelectric-point markers to accurately determine the pI of some proteins and a peptide by CIEF (Shimura *et al.,* 2000). Other studies reported a multidimensional system to separate tryptic peptides by CIEF coupled to nano-HPLC (Kuroda *et al.,* 2005) or capillary RPLC (Chen *et al.,* 2003). IEF separation of yeast cytosol peptides in fused silica capillaries was performed by Shen *et al.*, and resolved peptides as close as 0.01 pI units (Shen *et al,*. 2000). Thus, CIEF shows a high resolving power. However, the low loading capacity of this technique and the fact that the system assembly is composed of complicated switching valves capillary traps, makes the routine use of this system difficult, especially for non-expert users. Capillary electrophoresis application to peptides analysis has been extensively reviewed by Kasicka (Kasicka, 2008).

These limitations have driven researchers to use other techniques, which allow for larger sample volumes, but are still gel-free approaches.

### III.A.2. Solution IEF

Karger's group was amongst the first to use solution IEF of peptides, and showed the advantage of fractioning peptides into unique fractions generally not possible by cation-exchange chromatography (Tan *et al.,* 2002). They constructed a miniaturized multi-chamber device with a high sample load of 10 mg of peptides. Preparative Rotofor (Bio-Rad, Hercules, CA, USA) was also used by Groleau and colleagues to fractionate a solution of tryptic hydrolysate of bovine beta-lactoglobulin without ampholytes in order to identify interactions between peptides (Groleau *et al.,* 2002). Xiao *et al.* utilized the concept of "autofocusing", which was first presented by Yata and colleagues (Yata *et al.,* 1996) almost a decade before their study (Xiao *et al.,* 2004). They fractionated human serum peptides with a Rotofor in an ampholyte-free method. They identified 844 unique peptides that correspondied to 437 proteins. The authors argued that ampholyte-free IEF is useful to avoid purification steps that result in peptide loss. An and co-workers presented a two-dimensional strategy for peptide separation. This strategy incorporated solution IEF with off-line RPLC (An *et al.,* 2005). Insoluble nuclear proteins isolated from human MCF-7 breast cancer cells were used to evaluate their method. This 2D method was combined with $^{18}$O labels for differential analysis of protein expression between two cell lines. Yates's group also presented the use ZOOM® IEF fractionator (Zuo and Speicher 2000), which has 7 compartments separated by polyacrylamide discs in fritted polyethylene supports (Cantin *et al.* 2006). They combined IEF, immobilized metal affinity chromatography (IMAC) and LC-MS to enhance the detection of phosphorylation sites within complex protein samples.

Although solution IEF allows separation of peptides without the use of any detergent and polymeric reagent, free-flow electrophoresis (FFE) has the advantage of narrow-range separations (Moritz *et al.,* 2004). FFE was born as a technique to purify cells and subcellular organelles, which could be recovered highly purified as thin zones, due to their very low diffusion coefficients. FFE would not be ideal for protein prefractionation, though, due to their higher diffusion coefficients, as compared with cells and organelles. However, Kobayashi *et al.* reported a micro-fabricated FFE device that was useful for continuous separation of proteins (Kobayashi 2004). Baczek also used a home-made solution IEF device to focus standard protein and *Sacchromyce cerevisiae*-derived peptides (Baczek, 2004; Baczek, 2004). The Baczek study demonstrates one of the disadvantages of liquid-based IEF, which is its low resolving power. In their study, 38% of peptides were found in more than one fraction and 15% were found in more than four fractions. Continuous FFE device was also used to focus cultured human colon cell peptides by Simpson's group (Moritz *et al.,* 2004) and to analyze a human leukemia cell line by Wang and colleagues (Wang *et al.,* 2004). Griffin's group also reported the use of FFE to analyze shotgun-produced peptides of the chromatin-enriched fraction of *Sacchromyces cerevisiae* (Xie et al., 2005). They used peptide pI values as a filtering criterion to help to confirm the identification of some single-hit proteins and semi-tryptic peptides. These identifications were cross-validated with immunoblotting. Aebersold's group optimized the FFE protocol with a flat pI gradient and with the use of mannitol and urea in the separation media to simplify sample cleanup before downstream LC-MS/MS (Malmstrom *et al.,* 2006). Using samples from *Drosophila melanogaster*, they also incorporated the added value of pI in PeptideProphet software (Keller *et al.,* 2002) to improve data validation. Recently, Griffin's group used the added pI value of the peptides to enhance identification of PTM (Xie *et al.,* 2007). A peptide mixture from a yeast whole-cell lysate was fractionated with FFE and analyzed by LCMS/MS. They demonstrated the potential of FFE in proteomic studies of protein PTMs. The peptide pI prediction algorithm was adjusted to account for the elimination of charge on lysine residues by acetylation, and these shifted pI values were used to improve the confidence of identifying modified peptides. FFE has a high sample load, short separation time, and no regeneration time between consecutive runs: however, the gradient produced in liquid-based IEF technologies such as FFE is not as reproducible as the one immobilized in the IPG strips.

### III.A.3. IPG-IEF

Loo and co-workers developed a bottom-up approach called "virtual 2D" in which, after focusing, IPG strips were washed and sliced without detaching the gel from the plastic backing (Loo *et al.,* 2005). On-gel trypsin digestion was perfomed, and a-cyano-4-hydroxycinnamic acid (CHCA) was applied. Dried, matrix-impregnated strips were mounted on the MALDI sample stage, and proteins were analyzed. Although their approach allowed a direct analysis of the samples from the strip, the diffusion in sliced fractions and the inefficient digestion of the proteins reduced its capacity to provide high resolution. The contaminants from the strip such as the ampholytes, CHAPS, and urea also disturbed the signal. Virtual 2D inherited the main disadvantage of the 2-DE gels, which is its inability to analyze membrane and hydrophobic proteins.

In 2004, Loo's group published the first paper on the use of IPG as the first dimension in Shotgun proteomics (Cargile *et al,.* 2004). In this methodology, proteins were digested with trypsin and purified before being applied to IPG strips by an over-night rehydration method. After focusing, IPG strips were dissected into a number of fractions. Peptides were extracted from the fractions with using a series of washes in acidic solvents. Extracted peptides were separated in second dimension by RPLC. Following LC, peptides were eluted directly onto a MALDI target with a spotting robot or sprayed in an ESI instrument. This step was followed by data-dependent tandem mass spectrometry on a MALDI or an ESI instrument. Spectra were submitted to protein identification and data analysis with powerful bioinformatics tools. Because peptides are separated in the first dimension by their pI and in the second dimension on a RPLC column by their retention time (Rt), these two parameters could be used to coroborate MS identification results. Using IEF-RPLC-MS/MS, Cargile and co-workers could identify more than 6000 peptides from more than 1200 *Escherichia coli* proteins, from approximately 10µg of starting material. They also showed that this technique improved the sensitivity by 50-100% compared to ion-exchange chromatography (strong cation exchange, SCX). Using pI to filter their data on *E.coli* proteome analyzed by Shotgun IEF, they could adjust the SEQUEST cross-correlation score (Xcorr) to produce a low peptide false-positive rate of 1% whilst increasing by 23% the number of validated peptides.

Later, Heck's group compared various time and voltage running conditions for in-gel peptide IEF (Krijgsveld *et al.,* 2006). Optimal focusing conditions were considered to be reflected in a low standard deviation of the average pI and in a narrow interval between the most extreme pIs of the peptides found in one gel slice. They found that focusing occurs even with low voltages. Their data showed that IEF of peptides is a relatively fast process, with focusing times that are ca. 10-times shorter than for proteins. This speed is most likely due to the much smaller peptides that migrate more easily through the gel than the larger proteins. Proper focusing seemed to be achieved more easily in the acidic half compared to the basic half of the strip. However, peptides with a basic pI required longer running times and/or higher focusing voltages than acidic peptides so that the running conditions of an in-gel peptide IEF might directly depend on the chosen pH range of the IPG strip. In contrast to protein IEF, peptide IEF is more tolerant to salt. Although under such conditions high voltages are not reached, efficient focusing is still achieved.

Shotgun IPG-IEF has been recently applied to a broad range of biological applications, and was used to profile different strains of *Staphylococcus aureus* bacteria (Scherl et al., 2006). Several strains, with different susceptibility levels to vancomycin or teicoplanin, were subjected to whole-genome microarray-based transcription and quantitative proteomic profiling by combining iTRAQ with Shotgun IPG-IEF. Over 5000 unique peptides that represent more than 850 proteins were identified and quantified in a single experiment from the membrane fraction. This number represents more than 30% of the predicted *S. aureus* proteome. More than 250 peptides were identified in a single fraction. In addition, a series of genes and proteins that are possibly involved in the resistance mechanisms were discovered by the analysis of the quantitative data. In this experiment 19% of the identified proteins were predicted to be integral membrane proteins. Theoretically, 26% of the *S. aureus* proteins (668 proteins) are predicted to contain one or more Transmembrane (TM) segments. Membrane proteins are inherently difficult to solubilize and maintain in solution during digestion. Nevertheless, Scherl and co-workers showed that, by comparing the theoretical number of proteins with TM segments and proteins identified with GeLC-MS, shotgun IPG-IEF resulted in the highest number of identified membrane proteins. Lengqvist *et al.* also reported the use of iTRAQ compatibility with the Shotgun IPG-IEF technology by using standard proteins and total cytosolic protein fractions from a human colon cancer cell line (Lengqvist *et al,.* 2007). They reported that individual peptides with separate iTRAQ labels co-migrate in IEF, because the functional groups in all the tags are the same. They also found that iTRAQ labeling induces a small pI-shift compared to the native peptides that is negligible in the acidic pH range. With shotgun IPG-IEF, Chick *et al.* compared tryptic digestion of membrane proteins from rat liver, that was carried out in varying concentrations of methanol in 10 mM ammonium bicarbonate: 0%, 40%, and 60% (v/v). A total of 800 proteins identified from 60% methanol increased the protein identifications by 17% and 14% compared to 0% methanol and 40% methanol-assisted digestion (Chick *et al.,* 2008). Sevinsky and co-workers reported a modified protocol for $^{18}O/^{16}O$-labeling for shotgun IPG-IEF (Sevinsky *et al.,* 2007). They estimated a different protein expression in human plasma samples from a study on innate and adaptive immunity to infectious disease. By using immobilized trypsin in the initial digestion step, they minimized any post-labeling back-exchange of $^{18}O$-labeled peptides into $^{16}O$. Bunger *et al.* employed shotgun IPG-IEF to detect SNPs in breast tumor cells (Bunger *et al.,* 2007). Out of a total of 629 SNPs, 36 were of alternate SNP alleles not found in the reference NCBI or IPI protein databases. Their work highlights that the usefulness of interpreting unassigned spectra as polymorphisms is highly reliant on the ability to detect and filter false positives with pI. Bunger *et al.* demonstrated another use of pI filtering to identify coding non-synonymous SNPs in un-assigned MS/MS spectra.

A more widely used first step to separate complex mixtures has been SDS-PAGE gel electrophoresis (GeLC-MS). In a pair-wise comparative study on *Drosophila* nuclear extracts, using the same amount of starting material, Krijgsveld and co-workers compared the two technologies. Shotgun IPG-IEF resulted in a 27% increase in the number of unique peptide identifications with a FT MS instrument. With a LTQ MS, 16% more unique peptides were identified with IPG-IEF. This increased sensitivity of IEF relative to SDS-PAGE could be related to its resolving power (Krijgsveld *et al.,* 2006). On the peptide separation side, SCX has long been the 'gold standard' for multidimensional separation coupled to Shotgun approaches (Peng *et al.* 2003). In 2005, Essader and co-workers published a direct comparison between IEF and SCX as first dimension on shotgun proteomics for the analysis of testis from *Rattus norvegicus* (Essader *et al.,* 2005). Narrow-range IPG strips were used and identified 17% more peptides and 13% more proteins than an optimized off-line SCX approach. They also showed that the efficiency of the instrument time is much greater for narrow-range IPG than for the equivalent SCX experiment. The major drawback with IPG is that it currently takes more user time to prepare the samples than the equivalent SCX approach. With MS as the rate-limiting step, the method duty cycle is equivalent if multiple samples are analyzed. With IPG, more than one sample can be rehydrated and focused at once, provided that all samples have the same matrix. The Shotgun IPG-IEF method has several merits that make it a promising method for shotgun proteomics. These merits include high capacity and a wide dynamic range in combination with a high-resolution separation. However, the IPG gradients still need some carrier ampholytes to optimize separation, probably to smooth the conductivity differences existing along an immobilized pH gradient

(Righetti, 1983).

### III.A.4. Off-Gel

A system called "Off-Gel IEF" has been described by Ros and co-workers (Ros *et al.*, 2002). Just like the multi-compartment separation technique, the off-gel system has been devised to separate proteins and peptides according to their pI, and for their direct recovery in solution without any buffers or ampholytes. The sample is placed in a liquid chamber that is positioned on top of an IPG gel. Theoretical calculations and modeling have shown that the protonation of an ampholyte occurs in the thin layer of solution close to the IPG gel/solution interface. Upon application of a voltage gradient perpendicular to the liquid chamber, the electric field penetrates in the channel and extracts all charged species i.e., with pI value above and below the pH of the IPG gel, thus removing them from the sample cup. After separation, only the globally neutral species (zero net charge) remain in solution. Hörth and colleagues used Agilent's 3100 OFFGEL to fractionate *E.coli* peptides (Horth *et al.*, 2006). They identified 3454 peptides and 670 proteins, and found that 74% of the peptides were found in only one fraction. However 1/6 of their fractions contained fewer than 5 peptides. Heller presented a two-stage Off-Gel IEF of proteins and peptides (Heller *et al.* 2005). Intact human plasma protein isoforms that differed in pI were separated in a first-stage IEF followed by protein digestion and focusing of the derived peptides by the same device. A total of 15 protein and 15 peptide fractions yielded a very large scale (225 fractions) proteome analysis. After immunodepletion, they identified 660 peptides of 81 proteins of a single protein fraction in the pI zone 5.00 to 5.15. In another study, the same group separated peptides derived from a cellular extract of *Drosophila Kc167* cells in 15 fractions. They demonstrated the value of pI and Rt to validate peptides that were identified by two search engines (Heller *et al.*, 2005). Off-Gel has a high loading capacity, and has the major advantage of an in solution recovery of the peptides. However, the main disadvantage of the commercial Off-Gel devices is that their number of fractions is predefined, and the user is limited by the predefined shape and size of the wells. Additionally, in comparison to shotgun IPG-IEF, Off-Gel is more time-consuming. The main reason is that, in Off-Gel, the electrical field has vectors in two dimensions whereas in simple IPG-IEF the electrical field projects only in one direction. Consequently, IPG-IEF is a faster procedure than the Off-Gel.

Compared to other separation techniques, IEF has shown a superior resolution, because the focusing effect allows a very narrow peak width. The peak width is inversely related to the voltage, and most of the techniques described above (e.g., capillary IEF, and IPG-IEF) use very high voltages.

In a completely different but also very important trend, IEF relies on a parameter (pI) that can be easily calculated from the amino acid sequence of the peptide and/or protein, and this pI value offers interesting insights to corroborate the results obtained from the mass spectrometry analysis by taking into account the experimental pI of the peptide. This situation differs from what is encountered in chromatography, where retention times (either in ion exchange or in reverse phase) are more difficult to predict, because of the additional parameter represented by the nature of the static phase.

### III.A.5. A new dimension for data validation

The ability to calculate the pI of proteins within a defined pH gradient was first demonstrated by Bjellqvist and co-workers for 2D gel electrophoresis (Bjellqvist *et al.,* 1993). This approach is based on knowing the pK differences between closely related immobilines by focusing the same sample in overlapping pH gradients. A key component of this experiment is that the ratio of the immobilized groups incorporated into the gel matrix is similar to the ratio found in solution, so that a correct determination of the pK values can be obtained. Using this approach, protein pIs could be predicted within 0.2 pI unit precision. Other approaches have included the use of fluorescent markers in CIEF (Shimura *et al.* 2000; Shimura *et al.* 2002) or the use of pI and titration-curve calculations to predict pI shift for PTMs (Gordon *et al.,* 2005).

Shotgun IPG-IEF provides an increased dynamic range and experimental database filtering capabilities for false positive and negative identifications. Most pI calculation algorithms published to date are exclusively based on charged amino acid side chains in a peptide or protein (Bjellqvist *et al.,* 1993; Shimura *et al,.* 2000). Recently, Cargile and co-workers studied the effect of adjacent amino acid residues on peptide pI (Cargile submitted). They reported a computational method based on a genetic algorithm (Schmitt 2001). This method takes into account three consecutive amino acid residues and their cumulative effect on pI. For pK calculations on the C-terminus, their genetic function took into account any amino acid within three residues of the basic lysine (K) or arginine (R) moiety found in a tryptic peptide. A correction factor for aspartic acid (D) and glutamic acid (E) residues and their interaction with the N- or C-terminus is considered, provided that these amino acids are within four residues of that terminus. Their algorithm was trained on 5000 unique tryptic peptides. This method accurately calculates pI within a range of +/- 0.05 to +/- 0.03 pI units for the acidic part of the gradient. A comparison of the usual algorithm used for proteins and the new algorithm shows a much narrower distribution of peptides across a IPG strip's gradient as shown in Figure 3.. For the rest of the gradient, Bellqvist's algorithm was used with corrections for N- and C- terminal residues with charged side chains (Bjellqvist *et al.,* 1993). This algorithm was used for all pI calculations throughout this work.

One of the major issues in MS-based proteomics is the validation of protein and peptide identifications produced by search engines such as Phenyx (Colinge *et al.* , 2003), Mascot (Perkins *et al.,* 1999) and SEQUEST (Eng *et al,.* 1994). There are several validation strategies that are based on empirically determined rules such as using a score cutoff (Tabb *et al.,* 2002) or manual tandem mass spectra validation (Link *et al.,* 1999). These approaches lack any firm mathematical basis. Because in shotgun IPG-IEF, peptides are separated according to their pI, which is an invariant physicochemical property, this "added value" can be exploited to filter protein/peptide databases. Cargile and co-workers reported the use of pI as a filtering criterion, which resulted in the match of greater than 1300 protein loci identified by a single peptide match (Cargile *et al.* 2004). They also showed a 23% increase in the number of identified peptides with the pI filtering compared to the SEQUEST's $X_{corr}$ score alone in an *E.coli* database. Other groups that used shotgun Off-GEL IEF (Krijgsveld *et al.*, 2006) and shotgun FFE-IEF (Xie *et al.,* 2005) also used pI to recover more false negative peptides while maintaining a low false positive rate, and to confidently identify single hit proteins. Chick *et al.* evaluated a two-step sequential filtering approach that used false discovery rates (protFDR) at the protein level to assign high-confidence protein identifications (Chick *et al.* 2008). The distinction between protFDR was that it was calculated at the protein level, where protein scores were ranked and a threshold level was set beyond which assignments were incorrect. However, false positive rates (pepFPR) were calculated at the peptide level, where the following formula applied, 2n(reverse)/n(reverse) + n(forward), and where reverse referred to reverse database assignments and forward refers to forward database. Their results demonstrated that peptide pI filtering prior to imposing a 1% protFDR increased protein identifications by at least 19% compared to using only a 1% protFDR.

The use of pI in combination with other orthogonal values becomes even more efficient. Cargile *et al.* showed the utility of using pI in combination with accurate mass (Cargile and Stephenson, 2004). Petritis *et al.* introduced methods to partially predict the elution time of peptides from reverse-phase columns, and to use this information in the protein identification (Petritis *et al.*, 2003). This information allows some discrimination between peptides to be made, but the ± 10% standard deviation in elution time leaves a large window within which a peptide could theoretically elute. It was shown from offline cation exchange that there is a significant correlation between solution-phase charge and the general elution time of peptides (Peng *et al.*, 2003). Combination of Rt and pI would also be a powerful tool for data validation (Krokhin, 2006). Orthogonal peptide pI filtering is not limited to the identification results. Accurate peptide pI can be applied prior to database searching to constrain the database size and complexity. Sevinsky *et al.* presented GENQUEST (GENome Queries Using Experimental SpecTra), a method that uses accurate peptide pI to constrain the database size of a six-frame translated human genome, that resulted in accurate and sensitive genome searching comparable to searching the human protein databases (Sevinsky *et al.*, 2008).

## III.B. Electrophoretic separation of proteins.

As analytes, proteins are much more complex than peptides, but also much more amenable to separation. Just to keep with generic parameters, proteins differ widely in size, isoelectric point, and hydropathy, which are parameters that can be used for separation. As to electrophoretic separations, only size and isoelectric point are used.
However, one of the major problems linked to protein separation is their chemical complexity. Nucleic acids are also long and complex polymers, but their sugar phosphate backbone and the low chemical diversity of the bases gives them a uniform behavior. Conversely, the polypeptide backbone of proteins is proportionally much less important compared to the amino acids side chains, so that the physico-chemical behavior of proteins is extremely complex and strongly depends on the composition and sequence of every protein. In addition, this complexity leads to the 3D folding of proteins, which induces, in turn, a complex surface chemistry, because proteins show surface patches of various features (e.g. cationic, anionic or hydrophobic). Consequently, proteins have a strong tendency to stick to other surfaces, whether these surfaces are those of other cellular molecules or surfaces present in the separation processes.
Thus, when analyzing native proteins, two major problems occur:
i) the equilibrium for each protein between its free form and various complexes and
ii) a tendency to aggregate to other proteins in artefactual complexes and/or precipitates. This aggregation is especially true in the stacking dimension when discontinuous multiphasic gels are used (see section II)

### III.B.1. Electrophoretic separations of native proteins
For these reasons, electrophoretic separations of native proteins are very rarely used in proteomics. A happy exception to this rule is represented by the system called Blue-Native PAGE (Schägger and Von Jagow, 1991), where the proteins and protein supercomplexes are separated by electrophoresis at a neutral pH. To limit aggregation, a charge-shifting agent is added, i.e. a reagent that is able to stick to proteins without unfolding their structure while conferring to them a uniform electric charge. Thus, electrostatic repulsion is maximized between proteins to prevent aggregation. While this system has been designed initially for the analysis of mitochondrial protein complexes (Schägger and Von Jagow, 1991), it has also been applied either to dedicated studies on other types of complexes (e.g., nuclear complexes Nováková et al., 2006) or to even more general large-scale studies (Camacho-Carvajal et al., 2004).

However, despite this trick of the use of a charge-shifting agent, this type of separation is difficult to optimize, because several parameters must be optimized for each system (Schägger and Von Jagow, 1991). Moreover, even with an optimized separation, it is quite frequent to observe some trailing of the bands in BN-PAGE separations, that indicate insufficient protein solubility.

### III.B.2. Electrophoretic separations of denatured proteins: rationale

Because of this pivotal problem of protein solubility, the overwhelming majority of electrophoretic separations of proteins is made under denaturing conditions. Stricto sensu, denaturation just means the unfolding of proteins. Unfolding can be brought by physico-chemical perturbations, (e.g. heat or pH extremes), in which case the unfolded proteins aggregate together to form a massive precipitate. However, in some cases, denaturation is brought by conditions that unfold the proteins because they unfavor the internal protein forces in favor of protein-solvent interactions. In this case, proteins are made much less sticky, between themselves and toward surfaces, so that their solubility increases dramatically.

By increasing the solubility (low solubility is the major bottleneck in protein separations), and by decreasing the complexity of the system by dismantling protein complexes, denaturing electrophoresis of proteins is the method of choice for protein separation in proteomics studies.

This kind of solubilizing denaturation can be brought by two types of reagents. The first type is made of chaotropes; i.e., chemicals that disrupt the structure of water. In doing so, these chemicals alter the non-covalent bonds (hydrophobic interactions, hydrogen bonds) that structure proteins and thus unfold them. The chaotropic effect of many chemicals has been investigated (Gordon and Jencks, 1963), and from such studies, three efficient chaotropes can be used in protein chemistry: urea, thiourea, and guanidine. However, it must be kept in mind that, to be active, chaotropes must be used at multimolar concentrations, thereby precluding the use any ionic chaotrope in electrophoretic separations, and thus the use of guanidine, which is the most potent chaotrope. Because thiourea is not very soluble in water, and furthermore prevents acrylamide polymerization unless special systems are used (Rabilloud, 1998), the chaotrope of choice in electrophoretic separations is urea.

One of the main advantages of neutral chaotropes such as urea is that they do not alter either the molecular weight of the proteins or their isoelectric points, so that they are compatible with all types of electrophoretic methods.

However, chaotropes are still not the perfect solution. First of all, urea can undergo chemical reactions with protein amino groups, to lead to carbamylation (Cejka et al., 1968). This reaction alters the isoelectric point of the proteins, which is a problem in IEF. Furthermore, carbamylation (which is artefactual) and acetylation (which is a natural protein modification) induce almost the same mass shift on the peptides and proteins, so that an artefactuel carbamylation can be misleadingly taken for a bona fide biological acetylation.
Second, neutral chaotropes do increase protein solubility, but they just play on an increase in the water solubility of poorly soluble molecules. Thus, they cannot counteract completely protein aggregation, especially at high concentrations. Furthermore, neutral chaotropes have no action on electrostatic interactions. Thus, if such interactions can happen (especially at the low ionic strength conditions that prevail in IEF); then precipitation of proteins of opposite charge or artefacts in migration can occur (Wen et al., 1983).

The second type of solubilizing denaturing agents is composed of ionic detergents; the archetype is SDS (sodium dodecyl sulfate). Their mechanism of action is fairly different from the one of chaotropes. The efficient ionic detergents are made from a long and flexible hydrocarbon chain that is linked to an ionic, polar head. The detergent molecules will bind through their hydrophobic hydrocarbon tail to the hydrophobic amino acids. This binding favors amino acid-detergent interactions over amino acid-amino acid interactions, thereby promoting denaturation. Moreover, these protein-detergent hydrophobic interactions hold in place many detergent molecules on the proteins, thus anchoring many ionic charges of the same sign. This electrostatic repulsion further promotes denaturation, but also induces a very strong repulsion between protein molecules, and thus an extremely high protein solubility. However, this situation is achieved only when the proper balance between the hydrophilic and hydrophobic properties of the detergent is achieved. If the alkyl chain is too short, and the binding to the proteins is too weak, not enough detergent molecules are bound to promote optimal solubility. If the alkyl chain is too long, only a limited number of sites will exist on the proteins to accomodate such a long alkyl chain. The result is the same; i.e., not enough molecules fixed to ensure optimal solubility under electrophoretic conditions (Lopez et al. 1991). The proper balance differs from one class of detergent to another, but is close to 12 carbons for alkyl sulfate and sulfonates (Reynolds and Tanford, 1970) and of 16 carbons for quaternary ammoniums (MacFarlane, 1983).
Through this mechanism of action, properly chosen ionic detergents can solubilize almost all proteins to very high concentrations. The price to pay for this solubility is the complete disappearance of the protein native electric charge (or pI) that is masked by the charges brought by the detergent. It should also be stressed that the rationale for the efficiency of ionic detergents is hydrophobic interactions, so that concomitant use of chaotropes and detergents, also found in the literature, is not very helpful. As a matter of fact, urea has been used to remove SDS from proteins (Weber and Kuter, 1971).

### III.B.3. The implementation of denaturing protein electrophoresis in proteomics

The criticism that has been made of capillary electrophoresis in peptide separation holds true for proteins, and to an even greater extent. Because peptides are the final analytes in proteomics, minimal steps take place between the separation of peptides and their analysis in mass spectrometry. Proteins, conversely, are an indirect analyte for mass spectrometry in most cases, which means, in turn, that several steps must be carried out between protein separation and mass spectrometry. Because each of these steps has a specific, and sometimes low yield, being limited in protein amount, as is the case for capillary electrophoresis, precludes the use of this technique in proteomics.
Non-capillary free-flow electrophoresis is generally a rather low-resolution technique, and is therefore of low interest in proteomics studies. The only exception to this rule is IEF in free flow (Hoffmann et al., 2001), which shows a much larger capacity than conventional gel-based techniques. However, IEF of proteins has its own problems (detailed later) that does not make it a first choice for proteomics studies.

Thus, the vast majority of protein electrophoretic separations are made in gel media, most often in polyacrylamide gels. Besides the improved resolution, gel media offer the enormous advantage of a continuous separation, as well as additional advantages for downstream processing that will be detailed later (section IV).

**III.B.4. The workhorse of protein separation in proteomics: SDS gel electrophoresis**

In almost every proteomics scheme where a protein electrophoresis separation is carried out, SDS electrophoresis is encountered. This statement shows how widespread SDS electrophoresis is in proteomics. It can be used as the sole protein separation prior to digestion and peptide separation (e.g. in Lasonder et al., 2002, Schirle et al., 2003) but it can also be coupled with other protein separations, such as native electrophoresis (Schägger and Von Jagow, 1991), (Hisabori et al., 1991), ion exchange chromatography (Szponarski et al., 2004), denaturing electrophoresis (MacFarlane 1989, Rais et al., 2004), and of course isoelectric focusing, making the well-known high-resolution 2D electrophoresis.

Historically speaking, SDS electrophoresis in its modern form was born in 1970, when Laemmli (Laemmli, 1970) coupled the high resolution of the Tris chloride glycine system of Davis (Davis 1964) to the well-known denaturing power of SDS, which was already used in lower resolution continuous electrophoresis (Shapiro et al., 1967). This method immediately proved extremely versatile and thus became of extremely wide use.
This success can be explained by a combination of factors. The essential factor is the high affinity of SDS for proteins. As stated earlier, SDS is a happy compromise between hydrophobicity and hydrophilicity, and binds extremely well to almost all proteins. In addition to promoting protein denaturation, it confers a very strong negative charge to proteins, which masks their native charge. Consequently, the charge density becomes approximately constant for every protein, so that the separation is driven mainly by the molecular mass. Moreover, because the proteins do repel strongly each other, extremely high solubilities can be achieved. This high solubility allows one to take full advantage of the resolving power of the electrophoretic discontinuous systems, in which the high resolution is coupled with very high protein concentrations in the stacking gel. Although these high concentrations lead to protein precipitation, tailing, and loss of resolution in native systems, such problems are not encountered in the denaturing systems.

In such discontinuous SDS electrophoresis, the migration of the proteins is measured in the gel compared to the migration of the small ions front, visualized with a dye. Because of the extremely wide size distribution of proteins, it is often interesting to optimize this resolution according to the experimental needs. This optmization can be done by two mechanisms. In the first, and obvious one, the resolution is tuned by changing the pore size of the resolving gel, and thus the acrylamide concentration. When low percentage gels are used, the low molecular weight proteins migration will be almost unhampered by the gel, so that they will run with the small ion front. Meanwhile, high molecular weight proteins can enter more deeply into the gel, and are thus more separated each one from the other. Conversely, if the gel concentration is increased, then low molecular weight proteins will be unstacked from the small ions and separated each one from each other, whereas many high molecular weight proteins will be almost unable to enter the gel and will remain close to the top of the gel, with poor separation. A good compromise is, of course, to make a porosity gradient in the gel with an acrylamide concentration gradient, so that the resolution is optimized for all sizes of proteins. However, this approach is not without drawbacks. For example, low-concentration gels are very soft, while high-concentration gels are brittle, thereby limiting the scope of this way of optmizing resolution.

The other way to optmize the gel resolution is to play with the buffer system. As shown in buffer calculations (Jovin, 1973a), the speed of the moving ion boundary varies with the composition of the buffer system, in terms of the nature of the ions and in terms of pH. In anionic systems, such as the classical Tris-chloride-glycine system, the speed of the ions increases with increasing pH. Thus, decreasing the pH slows down the ions and makes the low-molecular weight proteins coalesce with the small ion front. However, by stretching the migration time, it also stretches the resolution of higher molecular weight proteins (Johnson, 1982). Conversely, increasing the pH will increase the mobility of the small ions, unstack the low molecular weight proteins, and therefore enable their analysis. However, by shortening the migration time, it will also decrease the resolution of the high molecular weight proteins.

Optimization of the buffer system is therefore another way of tuning the separation, which operates at constant gel concentration. Of course, combining pH changes and gel porosity changes is an optimal way of reaching an adequate resolution in extreme cases, such as giant proteins (Fritz et al., 1989).

However, this play on pH is limited in the case of the classical Tris-chloride-glycine system. in its standard setup, it operates at a gel pH of 8.8, which is already a bit too basic compared to the pK of Tris (8.05). Getting to an even more basic pH to analyze smaller proteins cannot be reached in a Tris system, but can be reached by a more basic system, based on ammediol (Bury, 1981), but at such a basic pH, the acrylamide gel starts to hydrolyze. In fact, It should be kept in mind that, in anionic systems, the real operating pH of the system is higher than the pH of the starting gel. For example, in the classical Davis system, a pH 8.8 gel operates at pH 9.5, according to Jovin's calculations.

Thus, a better way to analyze low-molecular weight proteins is to use buffer components with a pK lower than the one of glycine. Initial attempts have been made with MES (Kyte and Rodriguez, 1983), but better results have been obtained with tricine (Schägger and Von Jagow, 1987) or bicine (Wiltfang et al., 1991).

These findings led to a more general trend in the use of gels operating at a more acidic pH, to decrease the problems associated with alkaline hydrolysis of acrylamide. In addition, it was shown that Tricine-based gels show an even better resolution than glycine-based gels (Patton et al. 1991). However, in this case there is the opposite problem with Tris, because optimized resolution for 20-100kDa proteins is reached at pH 7, which again is too far from the buffering range of Tris.

Thus, even more flexibility can be offered by a system that operates close to the pK of Tris; e.g., a taurine-based system (Tastet et al., 2003). In this way, changing the pH of the buffer can be done either on the acidic side or on the basic side, to favor the resolution of high- or low-molecular weight proteins, respectively.

However, despite its outstanding capacities in terms of protein solubilization, versatility, and flexibility, SDS electrophoresis alone has too low a resolution to be used as the sole separation technique in proteomics. The only exception to this rule lies in interaction proteomics studies, where protein complexes isolated through a tag are first fished out of the cell lysate through a chromatographic process, and then analyzed (Rigaut et al., 1999). Most often, the complexity of the protein complex is low enough to be separable in individual proteins by a single SDS electrophoresis, which can resolve ca. 50 different protein species. It can be argued that, even in the case of low complexity samples, it can happen that two or more proteins have the same apparent molecular weight, which will cause them to comigrate, even with fine optimization of the separation. Should this comigration happen, some tricks can be used, such as changing the buffer system (Patton et al., 1991), replacing pure SDS by a mixture of alkyl sulfates (Lopez et al., 1991), or even replacing SDS by a mixture of alkyl sulfates and fatty alcohols, because this substitution can lead to the separation of proteins with extremely close molecular weight (Brown, 1988). This separation is due to the fact that these ancillary chemicals will bind to proteins in a structure- or sequence-dependent way, to induce a sequence-dependent electrophoretic mobility

In any event, when the sample to be analyzed becomes more complex, SDS electrophoresis is no longer sufficient in terms of separation, and must be coupled to other separation techniques. This additional separation can be made either before or after SDS electrophoresis. However, when rationalizing the possible choices, it must be kept in mind that SDS-denatured proteins are soluble in the presence of SDS, but will become easily insoluble when the SDS is removed. The few examples available in the literature demonstrate that interfacing SDS electrophoresis with another downstream protein separation is cumbersome (Siemankowski et al., 1978, Tuszynski et al., 1979, Nakamura et al., 1992). Thus, on a practical rule, when a complex separation scheme is needed, SDS electrophoresis is most often the last separation at the protein level. Other protein separations are carried out before SDS electrophoresis. whereas downstream separation is made at the peptide level, as in the GeLC method schematized on figue 4 (Lasonder et al., 2002, Schirle et al., 2003). This GeLC method, i.e., a combination of SDS electrophoresis of proteins, followed by in gel digestion and RP HPLC-MS/MS analysis of the resulting peptides, has shown the ability to deal with many types of samples, such as whole cell lysates (Lasonder et al., 2002) or subcellular fractions (Bell et al., 2001, Andersen et al., 2002). However, the method is still not perfect. As a matter of fact, even with the exquisite sensitivity of modern mass spectrometry, to be able to positively identify a protein implies to start from a few femtomoles. However, owing to the dynamic range of protein expression in most cells, having a few femtomoles of a minor protein will mean a fairly high total protein load, and consequently high concentrations of the most-abundant proteins (see for example the gel shown on Figure 4). Despite the high loading capacity of SDS electrophoresis, high protein loads often result in some spreading of the major protein peaks, at least just because of their Gaussian shape, not even to speak of tailing induced by overloading. This spreading means, in turn, that these major proteins are identified in many gel fractions, and can divert the MS/MS process and lead to miss the minor proteins. Complicated exclusion lists processes must then be implemented to alleviate this problem.

Thus, a higher resolution approach is to use another protein separation step before SDS electrophoresis.
Owing to the flexibility of SDS electrophoresis, almost any protein separation method, either chromatographic or electrophoretic, can be used prior to SDS electrophoresis. Among this wide variety, the coupling of gel-based electrophoretic separations has been most widely used, due to the fact that it represents a completely continuous bidimensional separation, without the sampling artefacts linked to the necessary sample collection when a liquid-separation is used as the first dimension.
Within this coupling of electrophoretic separations to SDS electrophoresis, the coupling of gel isoelectric focusing to SDS electrophoresis has received the widest audience.

### III.B.5. Classical high-resolution 2D electrophoresis: the source of proteomics

Coupling denaturing isoelectric focusing to SDS electrophoresis, as schematized on figure 5,

is almost ideal. Because the two methods use completely independent parameters, namely the isoelectric point and the molecular mass, the complete separation space will be used, without any skewing of the patterns. The first description of this two-dimensional method was made almost 35 years ago (MacGillivray and Rickwood, 1974). However, the IEF method was difficult to set in place (copolymerization of the sample within the gel with ampholytes), and the choice of the sample (nuclear proteins) and of the detection method (Coomassie blue) led to rather dull results. The following paper on the topic, made by Patrick O'Farrell (O'Farrell, 1975) had a much more important impact, because the results, obtained on radiolabelled E. coli, were very impressive, and the protocol was described in extensive detail, so that many researchers could make it work in their own laboratories. Because classical IEF with carrier ampholytes is prone to a major drawback called cathodic drift, that prevents the separation of basic proteins, an additional method was described shortly after to alleviate this problem (O'Farrell et al., 1977). At the end of the 70's, it was thus possible to separate in a relatively simple experiment several hundreds of polypeptides. It is, therefore, no surprise that many laboratories used this technique for different purposes, which can be classified first in cartographic purposes (Anderson et al., 1977), but also in differential purposes. In this case, the resolution of 2D electrophoresis was used to find out polypeptides whose amounts and/or synthesis rate changed between two physiological conditions, and just a few examples from the pioneers of that time are given (Garrels, 1979, Giometti et al., 1980, Bravo and Celis, 1980, Bravo and Celis, 1982).

To keep along a historical perspective, protein identification methods were in their infancy at that time, and it took several years of effort to identify a proliferation-associated polypeptide (Bravo et al. ,1981) as a subunit of a DNA polymerase (Bravo et al., 1987).

However, things evolved rapidly on this front of protein identification, first by Edman N-terminal microsequencing from blots (Matsudeira, 1987), then by internal Edman sequencing after peptide separation (Aebersold et al., 1987), and finally by mass spectrometry (James et al., 1993, Pappin et al., 1993, Yates et al., 1993, Henzel et al., 1993). Coupled with the development of immobilized pH gradients, which superseded the use of carrier ampholytes for IEF (Görg et al., 1988), this progress in protein identification was the birth of proteomics as we still know it now.

Besides this historical perspective, it is interesting to analyze the strengths and weaknesses of 2D electrophoresis in a modern proteomic approach.
Two-dimensional electrophoresis has a resolution that can be counted in thousands of polypeptides, especially when large-scale gels are used (Voris and Young, 1980, Young, 1984). This resolution is close to the expected complexity of many proteomes, and 2D gels have been used indeed to estimate this complexity (Duncan and McConkey, 1982). This situation is in sharp contrast with other two-dimensional methods, such as 2D liquid chormatographic separation of peptides. This method also has a resolution that can be above ten thousands of peptides, but the proteome complexity in that case is close to one order of magnitude higher in many cases.

Because of this very high resolution, polypeptide spots on a 2D gel often contain either one polypeptide, or a mixture of limited complexity (2-5 polypeptides) in which one protein accounts for most of the protein mass in the spot. This situation explains why a simple technique such as peptide mass fingerprinting can work on 2D gel spots, even with complex extracts, whereas it is bound to fail with SDS gels, in which one band contain many different proteins (see figure 6).

A second very important advantage of 2D gels in a modern proteomics setup is that it offers a very flexible and valuable platform to initially screen the proteins of interest, because 2D gels combine high resolution, low cost and image display of the results.

These features are best examplified (and used) in the case of differential proteomics, in which the goal of the experiment is to find the qualitative and more often quantitative changes at the proteome level between two or more samples. In this type of experiment, there are most often relatively few changes compared to the bulk of the proteome, which remains more or less constant.

In shotgun or GeLC techniques, the separation before the mass spectrometry offers no quantitative readout, so that all peptides, whether of interest or not, must be analyzed by the mass spectrometer. Such a process is very intensive in machine time, and is also generally difficult to parallelize. Moreover, this process generates an enormous flow of data, most of which are discarded because they belong to the constant part of the proteome. Thus, such heavy experiments are often not repeated with many biological replicates, with a concomitant decrease in the confidence that can be given to the output.

Oppositely, 2D gels offer a readable output as images before any mass spectrometry analysis is carried out. Thus, image analysis can be used first to select the proteins of interest, which will be submitted to digestion and mass spectrometry analysis. This process offers substantial savings in machine time. In other words, mass spectrometer time is exchanged against image analysis, which is a fairly efficient parallel process. Moreover, 2D gels are a simple experiment that can be repeated and parallelized easily. Thus, it is possible to produce enough images from enough samples to get statistical confidence at the 2D gels level; i.e., before spot selection and careful mass spectrometry analysis of the selected spots.

It is, therefore, not surprising that parallelization of the experiment (Anderson and Anderson, 1978a, Anderson and Anderson, 1978b) and computerized image analysis (Garrels, 1979, Anderson et al., 1981, Tarroux et al., 1987, Appel et al., 1988) have been implemented very early for 2D gels, even before proteins could be identified from gels. Important progress in the image analysis-based process was made with the use of immobilized pH gradients, which afford superior positional reproducibility of the spots (Corbett et al., 1994), and with the introduction of fluorescent multiplexing (Unlü et al., 1997), which allows several samples to be co-analyzed in a single gel, with minimal positional variation and, therefore, extremely precise detection of protein expression differences.

Another strong use of 2D gels is the work on post-translational modifications. IEF is exquisitely sensitive to changes in the pI of the proteins, and even on a large protein, a change of a single electric charge will produce a small shift in the pI, which is detectable on the 2D gel. This situation means that all modifications that add an electric charge (e.g. phosphorylation, sulfation, sialylation, deamidation) or remove one (e.g. N-acylation, amidation) will be detected, but that modifications that are electrically neutral (e.g., N-alkylation, S-acylation) will remain undetected.

For the modifications that can be detected, 2D electrophoresis has some important features:

i) it separates the modified form(s) from the bulk of the unmodified ones, which represents an enrichment for the modified peptides

ii) it keeps the filiation between the various protein forms: most often, the modification of pI brought by a single post-translational modification is very small, so that the unmodified protein can be found in close vicinity of the modified form(s). This proximity helps to make, in a second stage, a differential peptide analysis to assign the modification

iii) it makes no implicit hypothesis on the site of the modification. 2D gels can count the number of modifications but cannot locate them on the proteins. Taking the example of phosphorylations, 2D gels will separate mono-from di- from other multi-phosphorylated forms, but if several sites exist on the protein, protein mono-phosphorylated on site X will not be separated from protein mono-phosphorylated on site Y.

iv) it makes no implicit hypothesis on the nature of the modification. An acidic shift can mean whatever modification that adds a negative charge or removes a positive one.

Oppositely, searching for protein modifications directly at the peptide level is a completely different task. First, a modification on a peptide generally changes deeply its physico-chemical characteristics, which means that a modified peptide cannot be recognized from its parent peptide in a general peptide separation scheme. Thus, modified peptides must be selected on the basis of their modification. Such a selection is feasible for some modifications; e.g., phosphorylation (Hoffert et al., 2008) but much less for others; e.g., N-acylation or cysteine oxidation (Witze et al., 2007).

In other words, seeing a modified protein spot change on a 2D gel just "waves the flag" and indicates that some modification is taking place on the protein. Then, owing to the protein-separation power of 2D gels, it is possible to make a general search for the modified peptide(s) using mass spectrometry techniques, without prior selection of the modified peptides. This approach sometimes allows to find unsuspected but physiologically important modifications (Rabilloud et al., 2002), and also allows to map the site of modification, which cannot be done directly on the gel.

So, with such advantages and historical anteriority, how did it come that other proteomics methods could develop? This situation is just indicative of the weaknesses of 2D gels, which can be highly limiting in some cases.
The first important drawback is, in one sense, the other side of the coin, regarding the importance of image analysis in 2D gels. A single SDS electrophoresis lane can be systemically processed (cut, digest each slice, analyze the peptides), as is routinely done in the GeLC method. It is not feasible to cut systematically a 2D gel into thousands of gel cubes, and to process each of them. Thus, only what is visualized is analyzed in 2D gels. When dealing with complex proteomes (e.g., mammalian or plant cells or tissues), this situation means that low-abundance proteins are not analyzed at all, whereas there is always a probability for them to show up in a more systematic (shotgun or GeLC) analysis. Loading more material to make these low-abundance protein spots visible only reaches the stage where the gel is obscured by fused spots (see an example in Duncan and McConkey, 1982).

The second, and even more important, problem lies in the fact that some classes of proteins are very strongly under-represented on 2D gels, such as hydrophobic proteins and especially membrane proteins (Wilkins et al., 1998, Santoni et al., 2000), or extremely basic proteins. Although these examples are well known, other under-represented classes of proteins encompass proteins that are not solubilized from the cell in neutral chaotropes at low ionic strength; i.e., the conditions that prevail for the IEF step. Examples include some nuclear proteins (Gronow and Griffiths, 1971). Thus, 2D gels are generally unable to analyze efficiently nuclear proteins (Yuan et al., 2007), unless dedicated extraction/precipitation procedures are used (Henrich et al., 2007).

These weaknesses make 2D gels poorly adapted to cartographic proteomic purposes ("what is in sample X ? ") whereas they are nicely adapted to comparative proteomic purposes, as stated earlier. Moreover, it must be recalled that 2D gels are compatible with isotopic labeling, but are also a very efficient label-free quantitative proteomic tool by themselves.

However, the well-accepted intrinsic weakness of 2D gels toward membrane proteins has been well-characterized, and obviously lies in the IEF dimension (Eravci et al., 2008). This situation has prompted several authors to develop IEF-free 2D electrophoresis systems.

**III.B.6. Specialized, IEF-free 2D electrophoretic systems**

Practically speaking, electrophoresis is carried out according to two modes: isoelectric focusing and zone electrophoresis. Thus, an IEF-free 2D electrophoresis system will couple two types of zone electrophoresis. The case in which the first dimension is a native separation has been already mentioned (e.g., Schägger and Von Jagow, 1991) but systems in which both dimensions are denaturing have been used.
One of the oldest cases is represented by the specialized systems for the separation of ribosomal proteins (Mets and Bogorad, 1974), and is a good example of the general frame for such specialized 2D systems. As usual, the second dimension is a SDS electrophoresis, and the first dimension is another type of zone electrophoresis. Quite obviously, the separation in this first dimension should differ as much as possible from the separation obtained in the second dimension. Otherwise, the proteins would roughly lay on a diagonal, which represents a minor increase in resolution for a more complicated experimental setup.
It is therefore not surprising to see zone electrophoresis at low pH and in the presence of urea widely used in such 2D systems. Besides the system for ribosomal proteins, other systems of that type have been devised for histones (Hardison and Chalkley, 1978) and for nucleolar proteins (Orrick et al., 1973). However, such systems do not work properly for membrane proteins, which are not soluble enough in acidic urea in the presence of a nonionic detergent. In fact, proper solubilization of membrane proteins for electrophoretic fractionation requires an ionic detergent.

Thus, sticking to this rationale of a first dimension in an acidic system with an ionic detergent, the most obvious choice was the cationic/anionic system (MacFarlane 1989). A decisive step was the demonstration that this system was indeed able to solubilize and separate intrinsic membrane proteins (Hartinger et al., 1996). This finding prompted a wide use of this system for membrane proteomics (Langen et al., 2000, Navarre et al., 2002, Coughenour et al., 2004, Zahedi et al., 2005, Yamaguchi et al. 2008a, Yamaguchi et al. 2008b).

However, it has been recently shown (Burré et al., 2006) that some membrane proteins escape analysis by this system, probably because they are not solubilized, while being analyzed by the double SDS gel system (Rais et al., 2004). In this latter system, the off-diagonal effect is achieved through the use of urea in the first dimension, which alters the binding of SDS to proteins. In other systems, this altered mobility is achieved via a different buffer system in the two dimensions and high concentrations of glycerol (Williams et al., 2006).
Nevertheless, such double-zone systems with ionic detergents suffer from a lack of resolution. Especially in the high molecular weight zone, where the sieving effect of the gel dominates, the pattern is essentially diagonal, and the separation space opens when the molecular weight decreases.

Thus, there seems to be a kind of inverse correlation between resolution and solubilizing power. Even not to speak of the resolution of modified proteins, it seems obvious that the resolution of these systems is not sufficient to deal with the complexity of membrane preparations obtained from complex organisms. Thus, when compared, on the one hand, to classical 2D gels and, on the other hand, to GeLC methods, these double-zone 2D systems stand somewhat in the middle of the river. They rely on image patterns (and thus do not offer a rationale for a blind cutting of spots) while offering a low resolution and thus a high probability of piling several proteins under each spot.

Thus, obtaining the maximum coverage of a membrane subproteome might require a combinatorial approach: conventional, 2-DE ( separation of membrane-associated proteins or membrane proteins with minor hydrophobicity) and alternative gel-based methods such as GeLC, double SDS gel, 16-BAC/SDS or CTAB/SDS-PAGE (for highly hydrophobic membrane proteins) (Hunzinger et al. 2006, Braun et al. 2007).

## IV. From gels to peptides: additional features of electrophoresis in a proteomics strategy

The previous sections have dealt essentially with the separative aspects of electrophoresis. However, it is quite uncommon that the electrophoretic gels are used as such without any further processing. Most often, the proteins are digested to produce peptides, and the peptides are the analytes for the mass spectrometry. Quite often too; e.g., in strategies that rely on 2D gels, an image is produced from the gel, and this image is used to localize the proteins on the gels and also for quantitative purposes.

### IV.A. Detection of proteins after gel electrophoresis

Within this frame, important constraints lie on the protein detection method. It must be sensitive, linear, homogeneous (i.e., detect every protein with the same efficiency), and shall not disturb too much the downstream processes to be made on the protein (e.g., digestion or blotting). User-friendliness and low cost are also desirable features. Of course, no method fits this description, so that several detection methods can be used according to other experimental constraints, such as sample availability or equipment available in the laboratory.

The first family of methods is represented by organic dyes, and colloidal Coomassie blue has become a real standard in the field (Neuhoff et al., 1988). Although the method is simple, has a low cost, and offers superior compatibility with mass spectrometry, it lacks sensitivity, and the optimal sensitivity is reached only after a long staining time (Neuhoff et al., 1988). However, it is still very popular for the simple reason that any spot detected by this method contains more than enough material to perform a nice protein analysis through digestion and mass spectrometry of the peptides.

To alleviate this sensitivity problem, many different methods have been devised. Maybe the most original one is the zinc-imidazole staining (Fernandez-Patron et al., 1992), which relies in the delay of zinc precipitation in the gel induced by the presence of proteins. This method is more sensitive than Coomassie, is completed in less than one hour after electrophoresis, and is fully reversible. Because the only immobilization of proteins is induced by the presence of zinc, any zinc-chelating buffer will confer the proteins their original mobility. Consequently, this detection method interfaces wonderfully with protein blotting (Ortiz et al., 1992) or ex-gel elution, which are demanding methods, and also with in gel digestion (Castellanos-Serra et al., 1999). However, this method has been rarely used despite these obvious advantages for very simple reasons. First, the readout is fairly difficult, because proteins are translucent zones on a white background. Second, because of the precipitation-based mechanism, the linearity of the technique is very poor.

The next widely-used family of methods is represented by silver staining. Silver staining is almost as old as 2D electrophoresis (Switzer et al., 1979), and has evolved considerably with time. From the few principles at play in silver staining (Rabilloud, 1990), variations in almost all the steps have provided a host of protocols (probably more than 100 to date) of variable performances. Although modern silver-staining methods offer a good sensitivity (10-times better than colloidal Coomassie), a rather good homogeneity and a fair linearity (over one order of magnitude), coupled with simple labware and low cost, several caveats must be kept in mind with silver staining. On a practical point of view, all high performance silver staining methods require the use of solutions of limited stability. Consequently, most commercially available silver-staining kits do not perform as well-and by far- as "home-made" silver staining methods.

On a more fundamental point of view, proteins are very strongly immobilized in the gel after silver staining. This immobilization is due not only to the initial fixation of the proteins with acids and alcohols, as in the case of Coomassie Blue, but also to the crosslinking effect of formaldehyde, used as the image developing reagent (Richert et al, 2004). Consequently, it is almost impossible to elute proteins out of a silver-stained gel; e.g., for blotting (Wise and Lin, 1991), and the peptide yield is always lower for silver-stained gels compared to other methods such as Coomassie (Winkler et al, 2007), at equal loads of course. Moreover, it has been recently shown that other mass spectrometry artefacts are associated with the presence of formaldehyde (Osès-Prieto et al, 2007).

Nevertheless, silver staining is still a valid choice, especially when limited amounts of biological samples are available, because it allows protein detection, quantitation and identification via the in-gel digestion approach at protein amounts that are undetectable by other general detection methods. This feature has been recently examplified with new methods of silver staining, which do not use formaldehyde at all (Chevallet et al., 2008)

Because of the various limitations of these classical protein detection methods, a considerable amount of effort has been devoted to another family of methods, namely fluorescence-based detection. All fluorescence-based methods share very interesting features, such as an extremely good homogeneity of staining and a very wide linear dynamic range, most often over two orders of magnitude. However, these performances are reached only through the use of dedicated and expensive labware (laser fluorescence scanner). Fluorescence detection can be achieved via three different principles, namely covalent labelling, non-covalent affinity and environment-sensitive probes.

Covalent labelling is at play in the difference gel electrophoresis (Unlü et al., 1997). In this scheme, various reactive probes can be bound on the lysine or cysteine moieties of proteins, to allow sample multiplexing within a single gel. Very high sensitivities have been claimed for this scheme, but it should be kept in mind that, due to the limited number of labelling sites used, the real fluorescence signal for every protein is in fact very low. The high sensitivity claimed is obtained only though the use of very high performance fluorescence scanners, and to the fact that the background fluorescence is very low, thanks to the contrast afforded by the covalent binding scheme. With this covelant labeling scheme, spot excision for further identification cannot be achieved via this signal, but requires additional experiments, at least via another detection method.

Another important scheme for fluorescent protein staining relies on noncovalent adsorption of the fluorescent probe onto the proteins, quite similar to what is achieved by Coomassie Blue. In fact, Coomassie blue itself has also been used in that mode, taking advantage of its infrared fluorescence, with improved detection sensitivity compared to the visible absorbance (Luo et al., 2006). However, for practical reasons (e.g. spots excision for further analysis), visible fluorescence is preferred. The prototype of these protein-binding, fluorescent probes is represented by anionic ruthenium complexes (Berggren et al., 2000, Rabilloud et al., 2001). These compounds offer a much better sensitivity than Coomassie blue, close to silver staining, and better compatibility for downstream mass spectrometry analysis than silver staining (White et al., 2004), although this compatibility is still inferior to the one afforded by Coomassie Blue (Chevalier et al., 2004).

Another, even more versatile mode of fluorescent detection is achieved through the use of environment-sensitive probes. This generic term encompass several chemical families of compounds, which have in common the property to fluoresce strongly in hydrophobic environments but not in water. The first molecule used to achieve in-gel detection following this principle was Anilino Naphthalene sulfonate (ANS) (Daban and Aragay, 1984), with limited sensitivity. A much-improved sensitivity was achieved with other derivatives of the same family (Horowitz and Bowman, 1987) or with Nile red (Daban et al., 1991).
These methods offer an excellent versatility because of the mechanism of detection. When using protein-binding probes, which are generally anionic, the gels must be thoroughly fixed to remove all the SDS, and expose cationic proteins in a cationic solution, quite similarly to what is made for Coomassie Blue staining. This process is long but reproducible. Oppositely, in detection with environment-sensitive probes, advantage is taken of the existence of the SDS-protein complexes in the gel, because they provide the lipophilic environment in which the probes fluoresce. Thus, it is just necessary to disrupt the pure SDS micelles present in the protein-free zones of the gel to achieve an adequate contrast. Depending on the properties of the probes, this contrast can be achieved by minimal water rinses, as for the ANS derivatives or Nile Red. In some cases (Steinberg et al., 1996), the obtention of the contrast requires a harsher treatment with acetic acid. Alternatively, instead of relying on a dynamic removal of part of the SDS, the process can be completely controlled to ensure an optimal performance (Malone et al., 2001).
This flexibility in the gel treatment allows one to optimize the detection according to the constraints imposed by the downstream processes. When acid and/or alcohols are used, the situation is quite the same as with Coomassie Blue or ruthenium probes. However, when only water-based solutions are used for gel processing, the proteins can be mobilized out of the gels for downstream analyses (Steinberg et al., 2000). Quite often, one gel-processing type is linked to one class of fluorescent probes. However, it has been recently shown that carbocyanines can be used in the two schemes (fixing and non-fixing), with good results (Luche et al., 2007).

Protein detection with epicocconone (Mackintosh et al., 2003), marketed under the trade name of Deep Purple ®, relies probably on a mixture of all mechanisms, because the fluorescence properties of the molecules are environment dependent, whereas binding to the protein-SDS complexes and transient covalent binding to lysines through Schiff bases has also been suspected.

Last but certainly not least, protein detection by the means of radioactivity shall be mentioned. This technique was an important point in the demonstration of the power of 2D electrophoresis in the seminal paper of Patrick O'Farrell (O'Farrell, 1975) while previous work using the same combination of IEF and SDS PAGE but with protein detection by

Coomassie Blue was far more less impressive (MacGillivray and Rickwood, 1974). The sensitivity in the detection of radioisotopes was greatly increased first by the use of fluorography (Bonner and Laskey ; 1974) and later by the use of phsophor imaging technology (Johnston et al., 1990). Even today, detection of radioactive proteins is by far the most sensitive, homogeneous and linear process. However, to reach its maximal power, the radioisotope use shall have a short half life (to maximize the number of disintegrations in a given time) and an adequate radioactive energy (to facilitate detection on the support outside the gel, film or phosphor plate). This almost excludes 14C and 3H, i.e. the most versatile isotopes for protein chemistry, and 35S, 33P and 32P are the isotopes of choice for detection. While the phosphorus isotopes are highly valuable for workiug on phosphoproteins, 35S remains the sole isotope that can be used for high sensitivity and still high speed general protein detection. However, this means in turn that only metabolic labelling can be used with 35S-labelled amino acids. Thus, the use of radioactivity is almost exclusively confined to in vitro systems such as cell cultures, whereas detection methods using visible light (absorption or fluorescence) can be used on any type of sample. Finally, it must be underlined that working with radioactivity is more and more restricted in many countries, and that introducing radioactive peptides in a mass spectrometer is seldom welcome.

**IV.B. Production of peptides from gel-separated proteins**

It can be stated the age of proteomics started when the staining step stopped to be the last one of the process; i.e., when it became feasible to analyze the tiny amounts of proteins available from protein separation at a small scale, and especially through gel electrophoresis. For accuracy, sensitivity, and versatility reasons, protein microanalysis is now almost exclusively made through the use of mass spectrometry. Consequently, an important aspect of proteomics lies in the interface of protein separation with mass spectrometry, which is far from being trivial, because proteins, generally speaking, love what mass spectrometers hate: salts, detergents and polymers. Furthermore, when starting from a separation by gel electrophoresis, the protein is not freely accessible, because it is entangled in the gel matrix.
Except in very rare cases (Ogorzalek Loo et al., 1997), the proteins are not directly analyzed in the gel itself, so that three options are possible: i) the proteins are first eluted out from the gel, and processed for analysis, ii) the proteins are first transferred to a thin adsorptive phase (blotting), and processed for analysis, and iii) the proteins are digested into peptides in the gels, and the peptides are eluted out of the gel and analyzed.

Direct protein elution from gel pieces into a liquid phase is clearly not the favored option, although it has proven efficient in some cases (Cohen and Chait, 1997, Claverol et al., 2003). Proteins in an electrophesis gel are denatured and covered with SDS. If the SDS is removed, then the proteins have a strong tendency to precipitate, which is examplified when comparing the efficiency of water vs. SDS solution in protein elution from gels (Jørgensen et al., 2004).
Thus, a more efficient process consists in eluting the proteins from the gel and immobilizing them onto a thin adsorptive matrix. The idea of this process, also called blotting, is to make a replica of the gel on this matrix, thus keeping on the matrix the positional information obtained through the gel. The major advantage of this process is that the proteins are much more accessible on this matrix than in the gel, and this blotting process has gained enormous popularity first through immunoblotting (Towbin et al., 1979), but also through protein-chemistry methods such as Edman sequencing (Matsudeira, 1987). This blotting process offers several important advantages: i) it cleans up the proteins from the gel contaminants, ii) the proteins are more highly concentrated and more accessible than in a gel, iii) once blotted, the proteins cannot precipitate, so that the buffer can be changed at will, and iv) the blotting membrane itself can be stained to localize proteins, which eliminates the need for gel staining, with the additional advantage that membrane staining is faster and often more sensitive than gel staining (Sanchez et al., 1992). It is most often required to stain the blotting membrane only and not the gel before transfer, because all staining procedures that have a fixing step prevent protein transfer at least to some extent (Phelps, 1984, Wise and Lin, 1991), so that only a few gel-staining procedures can be used before protein transfer (Ortiz et al., 1992, Luche et al., 2007).

Because of these important advantages, it is not surprising that such a blotting step has been used as a handy interface between the gel and the mass spectrometer. In the most widely used mode of proteomics, peptides that derive from the digestion of the proteins of interest are the analytes in the mass spectrometer. Thus, important efforts have been devoted on how to digest efficiently proteins on blots, and also to recover the digestion peptides by elution (Lui et al., 1996, Courchesne, 1997, Van Oostveen et al., 1997, Bunai et al., 2003). Although showing some success, it is fair to say that these methods have not gained widespread use for a simple chemical reason. Just because the blotting membranes are made to adsorb proteins, they will also adsorb the proteases introduced to perform the digestion, and will adsorb the digestion peptides as well. Thus, antiadsorptive reagents must be used at this stage. They can be polymers, such as PVP40 (Henzel et al., 1993), or detergents (Lui et al., 1996, Van Oostveen et al., 1997), but both will induce problems in the mass spectrometer and/or in the reverse-phase separation of the peptides prior to mass spectrometry. Alternatively, high concentrations of acetonitrile have also been used (Bunai et al., 2003), but this procedure reduced the peptide yield.

In another mode of proteomics, the intact proteins are the analytes. In this case too, blots can be a good interface between the gel and the mass spectrometer. However, the general strategy for this type of analysis is not to desorb the proteins from the membrane, but to dissolve the membrane itself, which is possible when nitrocellulose membranes are used (Liang et al., 1996, Dukan et al., 1998, Luque-Garcia et al., 2006). However, here again, protein solubility after blotting is often an important issue.

Because of these built-in problems, the use of blots as an interface between electrophoresis and mass spectrometry is now rather confined to special cases in which the other, more classical methods do not work.
In fact, the enormous majority of the proteomics work is carried out by the third method, schematized on figure 7, in which the digestion is carried out in the gel pieces, followed by peptide extraction and mass spectrometry analysis of the eluted peptides. The feasibility of this in-gel digestion process had been proven before mass spectrometry was widely used for peptide analysis (Rosenfeld et al., 1992). Indeed, the needs of Edman sequencing (in peptide amounts and thus in peptide yields), were so demanding that it could be easily inferred that such a method, working well for Edman sequencing and thus reverse phase separation of peptides, would be perfectly suited for proteomics. However, here again, detergents had to be removed from all steps of the process (Reid et al., 1995). Although this removal is detrimental to the yield with blotting membranes due to their high adsorption abilities, it does not pose major problems in gels, due to their poor adsorption characteristics, which are selected for in gel electrophoresis.

The process has not widely changed since this time, and is still made of the same basic steps: i) dry the gel piece, ii) reswell it in a volatile buffer that contain the proteolytic enzyme (most often trypsin) and let digestion proceed, iii) collect the supernatant and extract the gel piece with a solution that contains a volatile acid and a solvent (most often acetonitrile), and iv) analyze the digest and extracts with mass spectrometry.

Alternatively, chemical cleavage methods can also be used (Tsugita et al., 2001), but the limited cleavage yields complicate the peptide map obtained for each protein because of frequent missed cleavages. Chemical cleavages are often preferred in case of membrane proteins for which the success of the identification is related to the number and length of hydrophilic loops that will produce peptides upon proteolysis. Indeed, transmembrane domains frequently lack of arginines and lysines and therefore, classical trypsin digestion yields mostly large, hydrophobic peptides.

It would be out of the scope of this paper to describe the variations made around this basic frame, especially in the peptide extraction solutions, which vary considerably in the solvent concentration as well as in the type and concentration of the acid. However, it is worth examining the real scope of the method, starting from the conditions that prevail during the process. In fact, the proteins are mobile after gel electrophoresis, but blotting and/or elution studies have shown that the proteins are mobile only as long as SDS is present on them, even in limited amounts. In all in-gel digestion processes, one of the first steps is gel washing with water/organic solvents (figure 7, top row). Such washes remove the gel buffer components, which would interfere with the downstream process, but also insolubilize the protein in the gel piece. Hence, the digested protein is not mobile, so that the protease must come into contact with the protein to be digested. This contact is achieved during the gel rehydration step, which is carried out in the presence of the protease (figure 7, middle row). In this step, the polacrylamide gel must be seen as a sponge that will absorb all components, provided that they are small enough to enter the gel network. This situation is, in fact, an important constraint on the proteases that can be used, but this often goes unnoticed because many useful proteases are small enough in size. Empirically, it appears that only proteases smaller than 25 kDa can be used in the in-gel digestion approach. Above this limit, the protease cannot enter efficiently in the gel, so that the digestion of the substrate is very limited and the extracted peptides are principally the protease autolysis peptides. This situation is encountered with some useful proteases, such as the Glu-C protease.
When this situation is encountered, approaches other than in-gel digestion should be used. One obvious choice is on-membrane digestion, as stated earlier, but another possibility is to implement the so-called peptide-mapping approach (Cleveland et al., 1977). In this approach, a piece of gel that contains the protein of interest is loaded together with the protease solution in the well of the SDS gel. During the initial phase of the electrophoresis, the substrate protein and the protease are coeluted and concentrated together in the stacking gel. By stopping the migration at this point, it is possible to carry out an in-gel digestion with almost any protease that is insensitive to 0.1% SDS, because the large pores of the stacking gel can accomodate a wider size range of proteases and substrates. Although this approach was created for limited digestion of proteins, it can be easily adapted to complete digestion: stop the gel during the stacking phase, dismantle it, excise the stacked zones, and let the digestion proceed in a tube.

This protease size problem illustrates a more general issue for in-gel digestion, which is the effective accessibility of the substrate protein in this process. Although the hundreds of publications that have used this approach have provided ample demonstration of the possibility to extract at least some peptides, some proteomics applications are more demanding than others and would ideally require a total coverage of the protein of interest. On a purely proteolytic point of view, coverage optimization requires one to develop conditions in which the substrate protein is denatured, to expose all possible cleavage sites, while the protease is still active. Fortunately enough, many proteases are quite resistant to denaturation, and work in low SDS concentrations, or rather high concentrations of urea or organic solvents.

From this starting point, the three main digestion schemes (in-solution, on-membrane and in-gel) can be compared in their yields and numbers of extracted peptides. This comparison has been carried out mainly between the in-solution and in-gel schemes. Rather surprisingly, inclusion of the sample in a gel seems to have an important beneficial effect on the overall peptide yield (Lu and Zhu, 2005, Han et al., 2008), which validates the power of the classical in-gel digestion approach. In-gel digestion is indeed a way of bringing the power and versatility of solid phase chemistry. It allows one to introduce the proteins in the gel in an optimally-denaturing solution, either by migration or by polymerization, to perform the digestion in conditions that are optimal for the enzyme, and to perform the extraction of peptides under optimized conditions again, whereas in-solution digestion is a single-step process in a single buffer, which is a compromise between denaturation and solubilization of the protein substrates, functioning of the enzyme, peptide solubility, and compatibility with further processes.

## V. Concluding remarks

Gel electrophoresis, and especially 2D gels, were quite important at the birth of proteomics. Indeed, proteomics arose when we could use protein microanalysis techniques, first Edman sequencing and then mass spectrometry, with the small amounts (at most a few micrograms) that could be separated by gel electrophoresis. To achieve such a result also implied to interface the two techniques, either by blotting or by using the gel piece itself.

With the development of proteomics, other techniques that do not rely on gel electrophoresis were invented (Washburn et al., 2001). Such approaches allowed to understand better the real strengths and weaknesses of gel electrophoresis. For example, it was shown that 2D gels were not suited for comprehensive proteomics, but still afford a superior screening capacity and a superior robustness, that thereby decreased the workload on mass spectrometers. The emerging trend is also that, when the adequate mode is used, (IEF for peptides and SDS PAGE for proteins) electrophoresis performs better than chromatography, although a final RP-nano-HPLC is always strongly needed just prior to the MS analysis.

Interestingly, comparative research has also shown the power of gels to interface protein separation with protein microanalysis, which could not be defined prior to the invention of fully gel-free techniques. Although gel-free approaches were initially pitched as replacements for 2DE-MS, it is likely that they will turn out to be complementary with their own limitations. In fact, as explained by Yates et al. (Lu et al. , 2008), one inherent disadvantage to the shotgun protein identification method such as MudPIT is the dependency on the acquisition of the MS/MS spectra. Yates has shown that for current MS technologies, it is impossible to perform MS/MS on every single ion of the LC-MS run. Kuster et al. estimate that less than 10 % of peptides presented to the mass spectrometer in adequate amounts actually contribute to useful peptide identification (Kuster et .al. , 2005). In order to overcome this limitation, Yates et al. have proposed to perform multiple MudPIT analyses (Liu et al. , 2004) or to use high mass accuracy mass spectrometers, such FT-ICR-MS and LTQ-Orbitrap (Lu et al; , 2008). Nevertheless, in these approaches described above, different drawbacks are encountered: 1. Storage and processing of huge MS data files , 2. Very long acquisition times, 3. Expensive MS instruments to generate high mass accuracy data necessary for efficient peptide identification, 4. Large dynamic range analyses, not compatible with most modern MS instruments, especially in the case of low abundance proteins. In this context, several authors have questioned the capacity of MS-based shotgun proteomics to ever determine a comprehensive coverage of any proteome using currently established methodology (Domon and Aebersold, 2006; Nielsen et al. , 2006 ; Cox and Mann, 2007). Nielsen et al. have considered that the extensive separation of proteins prior to digestion may be a viable alternative to the shotgun strategy, mainly for the detection of low abundance proteins (Nielsen, 2006). Today, we could conclude that 2DE-MS is now a mature and well established proteomics approach as reported by Figeys et al. : " 2DE-MS is now routinely used, while gel-free proteomic experiments are not and, conversely, are increasing in complexity" (Lambert et al. , 2005).

When integrating all the various features of the separation techniques and their interface with mass spectrometry, a decision tree can be made for proteomics experiments (see figure 8). In this figure, it can be seen that adequate electrophoresis techniques can be used with optimal benefit in almost all proteomics schemes. Of course, this decision tree is more flexible than it looks. For example, certain experimental designs require mandatorily to use quantification via stable isotopes, and in this case the cartographic scheme is better suited than the comparative one.

Within this general scheme, future areas for further improvement can be anticipated. A real high-resolution, IEF-free two-dimensional gel system is still to be invented, and would be well adapted to nuclear proteins and/or membrane proteins. Keeping with protein separations, isotachophoresis has probably not been investigated enough, because the self-implemented boundaries in isotachophoresis would decrease the cross-contaminations between proteins and thus increase the power of the GeLC method.

At the peptide level, and because of its incredible resolving power, IPG-IEF is definitely the method of choice. However, the practical use of this method can be further improved, to produce better overall performances by increasing the yield and efficiency of downstream processing of the peptides.

Thus, electrophoretic separations are clearly a major tool in proteomics strategies, and their use will probably further expand in the near future.

# VI. References

Aebersold RH, Leavitt J, Saavedra RA, Hood LE, Kent SB. 1987. Internal amino acid sequence analysis of proteins separated by one- or two-dimensional gel electrophoresis after in situ protease digestion on nitrocellulose. Proc Natl Acad Sci U S A 84:6970-6974.

An Y, Fu Z, Gutierrez P, Fenselau C. 2005. Solution isoelectric focusing for peptide analysis: comparative investigation of an insoluble nuclear protein fraction. J Proteome Res 4:2126-2132.

Andersen JS, Lyon CE, Fox AH, Leung AK, Lam YW, Steen H, Mann M, Lamond AI. 2002. Directed proteomic analysis of the human nucleolus. Curr Biol 12:1-11.

Anderson L, Anderson NG. 1977. High resolution two-dimensional electrophoresis of human plasma proteins. Proc Natl Acad Sci U S A 74:5421-5425.

Anderson NG, Anderson NL. 1978. Analytical techniques for cell fractions. XXI. Two-dimensional analysis of serum and tissue proteins: multiple isoelectric focusing. Anal Biochem 85:331-340.

Anderson NL, Anderson NG. 1978. Analytical techniques for cell fractions. XXII. Two-dimensional analysis of serum and tissue proteins: multiple gradient-slab gel electrophoresis. Anal Biochem 85:341-354.

Anderson NL, Anderson NG. 2002. The human plasma proteome: history, character, and diagnostic prospects. Mol Cell Proteomics 1:845-867.

Anderson NL, Taylor J, Scandora AE, Coulter BP, Anderson NG. 1981. The TYCHO system for computer analysis of two-dimensional gel electrophoresis patterns. Clin Chem 27:1807-1820.

Appel R, Hochstrasser D, Roch C, Funk M, Muller AF, Pellegrini C. 1988. Automatic classification of two-dimensional gel electrophoresis pictures by heuristic clustering analysis: a step toward machine learning. Electrophoresis 9:136-142.

Artoni G, Gianazza E, Zanoni M, Gelfi C, Tanzi MC, Barozzi C, Ferruti P, Righetti PG. 1984. Fractionation Techniques in a Hydro-Organic Environment .2. Acryloyl-Morpholine Polymers as a Matrix for Electrophoresis in Hydro-Organic Solvents. Analytical Biochemistry 137:420-428.

Baczek T. 2004. Fractionation of peptides and identification of proteins from Saccharomyces cerevisiae in proteomics with the use of reversed-phase capillary liquid chromatography and pI-based approach. J Pharm Biomed Anal 35:895-904.

Baczek T. 2004. Fractionation of peptides in proteomics with the use of pI-based approach and ZipTip pipette tips. J Pharm Biomed Anal 34:851-860.

Bell AW, Ward MA, Blackstock WP, Freeman HN, Choudhary JS, Lewis AP, Chotai D, Fazel A, Gushue JN, Paiement J, Palcy S, Chevet E, Lafreniere-Roula M, Solari R, Thomas DY, Rowley A, Bergeron JJ. 2001. Proteomics characterization of abundant Golgi membrane proteins. J Biol Chem 276:5152-5165.

Berggren K, Chernokalskaya E, Steinberg TH, Kemper C, Lopez MF, Diwu Z, Haugland RP, Patton WF. 2000. Background-free, high sensitivity staining of proteins in one- and two-dimensional sodium dodecyl sulfate-polyacrylamide gels using a luminescent ruthenium complex. Electrophoresis 21:2509-2521.

Bjellqvist B, Hughes GJ, Pasquali C, Paquet N, Ravier F, Sanchez JC, Frutiger S, Hochstrasser D. 1993. The focusing positions of polypeptides in immobilized pH gradients can be predicted from their amino acid sequences. Electrophoresis 14:1023-1031.

Bjellqvist B, Sanchez JC, Pasquali C, Ravier F, Paquet N, Frutiger S, Hughes GJ, Hochstrasser D. 1993. Micropreparative 2-Dimensional Electrophoresis Allowing the Separation of Samples Containing Milligram Amounts of Proteins. Electrophoresis 14:1375-1378.

Bonner WM, Laskey RA 1974. A film detection method for tritium-labelled proteins and nucleic acids in polyacrylamide gels. Eur J Biochem. 46:83-88

Braun RJ., Kinkl N., Beer M., Ueffing M. 2007 Two-dimensional electrophoresis of membrane proteins Anal. Bioanal. Chem.,389 : 1033-1045

Bravo R, Celis JE. 1980. A search for differential polypeptide synthesis throughout the cell cycle of HeLa cells. J Cell Biol 84:795-802.
Bravo R, Celis JE. 1982. Human proteins sensitive to neoplastic transformation in cultured epithelial and fibroblast cells. Clin Chem 28:949-954.
Bravo R, Fey SJ, Bellatin J, Larsen PM, Arevalo J, Celis JE. 1981. Identification of a nuclear and of a cytoplasmic polypeptide whose relative proportions are sensitive to changes in the rate of cell proliferation. Exp Cell Res 136:311-319.
Bravo R, Frank R, Blundell PA, Macdonald-Bravo H. 1987. Cyclin/PCNA is the auxiliary protein of DNA polymerase-delta. Nature 326:515-517.
Brown EG. 1988. Mixed anionic detergent/aliphatic alcohol-polyacrylamide gel electrophoresis alters the separation of proteins relative to conventional sodium dodecyl sulfate-polyacrylamide gel electrophoresis. Anal Biochem 174:337-348.
Bunai K, Nozaki M, Hamano M, Ogane S, Inoue T, Nemoto T, Nakanishi H, Yamane K. 2003. Proteomic analysis of acrylamide gel separated proteins immobilized on polyvinylidene difluoride membranes following proteolytic digestion in the presence of 80% acetonitrile. Proteomics 3:1738-1749.
Bunger MK, Cargile BJ, Sevinsky JR, Deyanova E, Yates NA, Hendrickson RC, Stephenson JL, Jr. 2007. Detection and validation of non-synonymous coding SNPs from orthogonal analysis of shotgun proteomics data. J Proteome Res 6:2331-2340.
Burre J, Beckhaus T, Schagger H, Corvey C, Hofmann S, Karas M, Zimmermann H, Volknandt W. 2006. Analysis of the synaptic vesicle proteome using three gel-based protein separation techniques. Proteomics 6:6250-6262.
Bury AF. 1981. Analysis of Protein and Peptide Mixtures - Evaluation of 3 Sodium Dodecyl Sulfate-Polyacrylamide Gel-Electrophoresis Buffer Systems. Journal of Chromatography 213:491-500.
Caglio S, Chiari M, Righetti PG. 1994. Get Polymerization in Detergents - Conversion Efficiency of Methylene-Blue Vs Persulfate Catalysis, as Investigated by Capillary Zone Electrophoresis. Electrophoresis 15:209-214.
Caglio S, Righetti PG. 1993. On the Efficiency of Methylene-Blue Versus Persulfate Catalysis of Polyacrylamide Gels, as Investigated by Capillary Zone Electrophoresis. Electrophoresis 14:997-1003.
Camacho-Carvajal MM, Wollscheid B, Aebersold R, Steimle V, Schamel WWA. 2004. Two-dimensional blue native/SDS gel electrophoresis of multi-protein complexes from whole cellular lysates - A proteomics approach. Molecular & Cellular Proteomics 3:176-182.
Cantin GT, Venable JD, Cociorva D, Yates JR, 3rd. 2006. Quantitative phosphoproteomic analysis of the tumor necrosis factor pathway. J Proteome Res 5:127-134.
Cargile BJ, Stephenson JL, Jr. 2004. An alternative to tandem mass spectrometry: isoelectric point and accurate mass for the identification of peptides. Anal Chem 76:267-275.
Cargile BJ, Talley DL, Stephenson JL, Jr. 2004. Immobilized pH gradients as a first dimension in shotgun proteomics and analysis of the accuracy of pI predictability of peptides. Electrophoresis 25:936-945.
Castellanos-Serra L, Proenza W, Huerta V, Moritz RL, Simpson RJ. 1999. Proteome analysis of polyacrylamide gel-separated proteins visualized by reversible negative staining using imidazole-zinc salts. Electrophoresis 20:732-737.
Cejka J, Vodrazka Z, Salak J. 1968. Carbamylation of globin in electrophoresis and chromatography in the presence of urea. Biochim Biophys Acta 154:589-591.
Chen J, Gao J, Lee CS. 2003. Dynamic enhancements of sample loading and analyte concentration in capillary isoelectric focusing for proteome studies. J Proteome Res 2:249-254.
Chevalier F, Rofidal V, Vanova P, Bergoin A, Rossignol M. 2004. Proteomic capacity of recent fluorescent dyes for protein staining. Phytochemistry 65:1499-1506.
Chiari M, Micheletti C, Nesi M, Fazio M, Righetti PG. 1994. Towards New Formulations for Polyacrylamide Matrices - N-Acryloylaminoethoxyethanol, a Novel Monomer Combining High

Hydrophilicity with Extreme Hydrolytic Stability. Electrophoresis 15:177-186.
Chiari M, Righetti PG, Negri A, Ceciliani F, Ronchi S. 1992. Preincubation with Cysteine Prevents Modification of Sulfhydryl-Groups in Proteins by Unreacted Acrylamide in a Gel. Electrophoresis 13:882-884.
Chick JM, Haynes PA, Molloy MP, Bjellqvist B, Baker MS, Len AC. 2008. Characterization of the rat liver membrane proteome using peptide immobilized pH gradient isoelectric focusing. J Proteome Res 7:1036-1045.
Claverol S, Burlet-Schiltz O, Gairin JE, Monsarrat B. 2003. Characterization of protein variants and post-translational modifications: ESI-MSn analyses of intact proteins eluted from polyacrylamide gels. Mol Cell Proteomics 2:483-493.
Cleveland DW, Fischer SG, Kirschner MW, Laemmli UK. 1977. Peptide mapping by limited proteolysis in sodium dodecyl sulfate and analysis by gel electrophoresis. J Biol Chem 252:1102-1106.
Cohen SL, Chait BT. 1997. Mass spectrometry of whole proteins eluted from sodium dodecyl sulfate-polyacrylamide gel electrophoresis gels. Anal Biochem 247:257-267.
Colinge J, Masselot A, Giron M, Dessingy T, Magnin J. 2003. OLAV: towards high-throughput tandem mass spectrometry data identification. Proteomics 3:1454-1463.
Corbett JM, Dunn MJ, Posch A, Gorg A. 1994. Positional reproducibility of protein spots in two-dimensional polyacrylamide gel electrophoresis using immobilised pH gradient isoelectric focusing in the first dimension: an interlaboratory comparison. Electrophoresis 15:1205-1211.
Coughenour HD, Spaulding RS, Thompson CM. 2004. The synaptic vesicle proteome: a comparative study in membrane protein identification. Proteomics 4:3141-3155.
Courchesne PL, Luethy R, Patterson SD. 1997. Comparison of in-gel and on-membrane digestion methods at low to sub-pmol level for subsequent peptide and fragment-ion mass analysis using matrix-assisted laser-desorption/ionization mass spectrometry. Electrophoresis 18:369-381.
Coute Y, Hernandez C, Appel RD, Sanchez JC, Margolles A. 2007. Labeling of Bifidobacterium longum cells with 13C-substituted leucine for quantitative proteomic analyses. Appl Environ Microbiol 73:5653-5656
Cox J, Mann M. 2007 Is proteomics the new genomics ?
Cell 130: 395-398..
Daban JR, Aragay AM. 1984. Rapid fluorescent staining of histones in sodium dodecyl sulfate-polyacrylamide gels. Anal Biochem 138:223-228.
Daban JR, Bartolome S, Samso M. 1991. Use of the hydrophobic probe Nile red for the fluorescent staining of protein bands in sodium dodecyl sulfate-polyacrylamide gels. Anal Biochem 199:169-174.
Davis BJ. 1964. Disc Electrophoresis. Ii. Method and Application to Human Serum Proteins. Ann N Y Acad Sci 121:404-427.
Domon B, Aebersold R. 2006 Challenges and opportunities in proteomics data analysis. Mol. Cell. Proteomics,  5: 1921-1926
Dukan S, Turlin E, Biville F, Bolbach G, Touati D, Tabet JC, Blais JC. 1998. Coupling 2D SDS-PAGE with CNBr cleavage and MALDI-TOFMS: a strategy applied to the identification of proteins induced by a hypochlorous acid stress in Escherichia coli. Anal Chem 70:4433-4440.
Duncan R, McConkey EH. 1982. How many proteins are there in a typical mammalian cell? Clin Chem 28:749-755.
Egen NB, Bliss M, Mayersohn M, Owens SM, Arnold L, Bier M. 1988. Isolation of monoclonal antibodies to phencyclidine from ascites fluid by preparative isoelectric focusing in the Rotofor. Anal Biochem 172:488-494.
Eng JK, McCormack AL, Yates JR. 1994. An Approach to Correlate Tandem Mass-Spectral Data of Peptides with Amino-Acid-Sequences in a Protein Database. Journal of the American Society for Mass Spectrometry 5:976-989.

Eravci M, Fuxius S, Broedel O, Weist S, Krause E, Stephanowitz H, Schluter H, Eravci S, Baumgartner A. 2008. The whereabouts of transmembrane proteins from rat brain synaptosomes during two-dimensional gel electrophoresis. Proteomics 8:1762-1770.
Essader AS, Cargile BJ, Bundy JL, Stephenson JL, Jr. 2005. A comparison of immobilized pH gradient isoelectric focusing and strong-cation-exchange chromatography as a first dimension in shotgun proteomics. Proteomics 5:24-34.
Fernandez-Patron C, Castellanos-Serra L, Rodriguez P. 1992. Reverse staining of sodium dodecyl sulfate polyacrylamide gels by imidazole-zinc salts: sensitive detection of unmodified proteins. Biotechniques 12:564-573.
Figeys D, Zhang Y, Aebersold R. 1998. Optimization of solid phase microextraction - capillary zone electrophoresis - mass spectrometry for high sensitivity protein identification. Electrophoresis 19:2338-2347.
Fritz JD, Swartz DR, Greaser ML. 1989. Factors affecting polyacrylamide gel electrophoresis and electroblotting of high-molecular-weight myofibrillar proteins. Anal Biochem 180:205-210.
Garrels JI. 1979. Changes in protein synthesis during myogenesis in a clonal cell line. Dev Biol 73:134-152.
Garrels JI. 1979. Two dimensional gel electrophoresis and computer analysis of proteins synthesized by clonal cell lines. J Biol Chem 254:7961-7977.
Gelfi C, Debesi P, Alloni A, Righetti PG, Lyubimova T, Briskman VA. 1992. Kinetics of Acrylamide Photopolymerization as Investigated by Capillary Zone Electrophoresis. Journal of Chromatography 598:277-285.
Gelfi C, Righetti PG. 1981. Polymerization Kinetics of Polyacrylamide Gels .2. Effect of Temperature. Electrophoresis 2:220-228.
Gelfi C, Righetti PG. 1981. Polymerization Kinetics of Polyacrylamide Gels .1. Effect of Different Cross-Linkers. Electrophoresis 2:213-219.
Giometti CS, Barany M, Danon MJ, Anderson NG. 1980. Muscle protein analysis. II. Two-dimensional electrophoresis of normal and diseased human skeletal muscle. Clin Chem 26:1152-1155.
Gordon JA, Jencks WP. 1963. The relationship of structure to the effectiveness of denaturing agents for proteins. Biochemistry 2:47-57.
Gordon JC, Myers JB, Folta T, Shoja V, Heath LS, Onufriev A. 2005. H++: a server for estimating pKas and adding missing hydrogens to macromolecules. Nucleic Acids Res 33:W368-371.
Gorg A, Boguth G, Kopf A, Reil G, Parlar H, Weiss W. 2002. Sample prefractionation with Sephadex isoelectric focusing prior to narrow pH range two-dimensional gels. Proteomics 2:1652-1657.
Gorg A, Obermaier C, Boguth G, Harder A, Scheibe B, Wildgruber R, Weiss W. 2000. The current state of two-dimensional electrophoresis with immobilized pH gradients. Electrophoresis 21:1037-1053.
Gorg A, Postel W, Gunther S. 1988. The current state of two-dimensional electrophoresis with immobilized pH gradients. Electrophoresis 9:531-546.
Graham RL, Sharma MK, Ternan NG, Weatherly DB, Tarleton RL, McMullan G. 2007. A semi-quantitative GeLC-MS analysis of temporal proteome expression in the emerging nosocomial pathogen Ochrobactrum anthropi. Genome Biol 8:R110.
Groleau PE, Jimenez-Flores R, Gauthier SF, Pouliot Y. 2002. Fractionation of beta-lactoglobulin tryptic peptides by ampholyte-free isoelectric focusing. J Agric Food Chem 50:578-583.
Gronow M, Griffiths G. 1971. Rapid isolation and separation of the non-histone proteins of rat liver nuclei. FEBS Lett 15:340-344.
Han CL, Chien CW, Chen WC, Chen YR, Wu CP, Li H, Chen YJ. 2008. A multiplexed quantitative strategy for membrane proteomics: Opportunities for mining therapeutic targets for autosomal-dominant polycystic kidney disease. Mol Cell Proteomics.in press

Hardison R, Chalkley R. 1978. Polyacrylamide gel electrophoretic fractionation of histones. Methods Cell Biol 17:235-251.

Hartinger J, Stenius K, Hogemann D, Jahn R. 1996. 16-BAC/SDS-PAGE: a two-dimensional gel electrophoresis system suitable for the separation of integral membrane proteins. Anal Biochem 240:126-133.

Heller M, Ye M, Michel PE, Morier P, Stalder D, Junger MA, Aebersold R, Reymond F, Rossier JS. 2005. Added value for tandem mass spectrometry shotgun proteomics data validation through isoelectric focusing of peptides. J Proteome Res 4:2273-2282.

Henrich S, Cordwell SJ, Crossett B, Baker MS, Christopherson RI. 2007. The nuclear proteome and DNA-binding fraction of human Raji lymphoma cells. Biochim Biophys Acta 1774:413-432.

Henzel WJ, Billeci TM, Stults JT, Wong SC, Grimley C, Watanabe C. 1993. Identifying proteins from two-dimensional gels by molecular mass searching of peptide fragments in protein sequence databases. Proc Natl Acad Sci U S A 90:5011-5015.

Hisabori T, Inoue K, Akabane Y, Iwakami S, Manabe K. 1991. Two-dimensional gel electrophoresis of the membrane-bound protein complexes, including photosystem I, of thylakoid membranes in the presence of sodium oligooxyethylene alkyl ether sulfate/dimethyl dodecylamine oxide and sodium dodecyl sulfate. J Biochem Biophys Methods 22:253-260.

Hoffert JD, Knepper MA. 2008. Taking aim at shotgun phosphoproteomics. Analytical Biochemistry 375:1-10.

Hoffmann P, Ji H, Moritz RL, Connolly LM, Frecklington DF, Layton MJ, Eddes JS, Simpson RJ. 2001. Continuous free-flow electrophoresis separation of cytosolic proteins from the human colon carcinoma cell line LIM 1215: a non two-dimensional gel electrophoresis-based proteome analysis strategy. Proteomics 1:807-818.

Horowitz PM, Bowman S. 1987. Ion-enhanced fluorescence staining of sodium dodecyl sulfate-polyacrylamide gels using bis(8-p-toluidino-1-naphthalenesulfonate). Anal Biochem 165:430-434.

Horth P, Miller CA, Preckel T, Wenz C. 2006. Efficient fractionation and improved protein identification by peptide OFFGEL electrophoresis. Mol Cell Proteomics 5:1968-1974.

Hunzinger C., Wozny W., Schwall GP., Poznanovic S., Stegmann W., Zengerling H., Schoepf R., Groebe K., Cahill MA., Osiewacz HD., Jaegemann N., Bloch M., Dencher NA., Krause F., Schrattenholz A.2006 Comparative profiling of the mammalian mitochondrial proteome: multiple aconitase-2 iosorms including N-formylkynurenine modifications as part of a protein biomarker signature for reactive oxidative species. J. Proteome Res. 5 : 625-633

James P, Quadroni M, Carafoli E, Gonnet G. 1993. Protein identification by mass profile fingerprinting. Biochem Biophys Res Commun 195:58-64.

Janini GM, Conrads TP, Veenstra TD, Issaq HJ. 2003. Development of a two-dimensional protein-peptide separation protocol for comprehensive proteome measurements. J Chromatogr B Analyt Technol Biomed Life Sci 787:43-51.

Johnson BF. 1982. Enhanced resolution in two-dimensional electrophoresis of low-molecular-weight proteins while utilizing enlarged gels. Anal Biochem 127:235-246.

Johnston RF, Pickett SC, Barker DL 1990. Autoradiography using storage phosphor technology. Electrophoresis 11:355-360.

Jorgensen CS, Jagd M, Sorensen BK, McGuire J, Barkholt V, Hojrup P, Houen G. 2004. Efficacy and compatibility with mass spectrometry of methods for elution of proteins from sodium dodecyl sulfate-polyacrylamide gels and polyvinyldifluoride membranes. Anal Biochem 330:87-97.

Jovin TM. 1973. Multiphasic zone electrophoresis. I. Steady-state moving-boundary systems formed by different electrolyte combinations. Biochemistry 12:871-879.

Jovin TM. 1973. Multiphasic zone electrophoresis. II. Design of integrated discontinuous buffer systems for analytical and preparative fractionation. Biochemistry 12:879-890.

Kasicka V. 2008. Recent developments in CE and CEC of peptides. Electrophoresis 29:179-206.

Keller A, Nesvizhskii AI, Kolker E, Aebersold R. 2002. Empirical statistical model to estimate the accuracy of peptide identifications made by MS/MS and database search. Anal Chem 74:5383-5392.
Kobayashi N. 2004. [Recent progress of free-flow electrophoresis method and its application for proteomics]. Tanpakushitsu Kakusan Koso 49:1333-1340.
Krijgsveld J, Gauci S, Dormeyer W, Heck AJ. 2006. In-gel isoelectric focusing of peptides as a tool for improved protein identification. J Proteome Res 5:1721-1730.
Krokhin OV. 2006. Sequence-specific retention calculator. Algorithm for peptide retention prediction in ion-pair RP-HPLC: application to 300- and 100-A pore size C18 sorbents. Anal Chem 78:7785-7795.
Kuroda Y, Yukinaga H, Kitano M, Noguchi T, Nemati M, Shibukawa A, Nakagawa T, Matsuzaki K. 2005. On-line capillary isoelectric focusing-mass spectrometry for quantitative analysis of peptides and proteins. J Pharm Biomed Anal 37:423-428.
Kuster B, Schirle M, Mallick P, Aebersold R. 2005 Scoring proteomes with proteotypic peptide probes Nat. Rev. Mol. Cell Biol., 6: 577-583
Kyte J, Rodriguez H. 1983. A discontinuous electrophoretic system for separating peptides on polyacrylamide gels. Anal Biochem 133:515-522.
Laemmli UK. 1970. Cleavage of structural proteins during the assembly of the head of bacteriophage T4. Nature 227:680-685.
Lambert JP, Ethier M, Smith JC, Figeys D.2005 Proteomics : from gel-based to gel free. Anal. Chem. 77: 3771-3788
Langen H, Takacs B, Evers S, Berndt P, Lahm HW, Wipf B, Gray C, Fountoulakis M. 2000. Two-dimensional map of the proteome of Haemophilus influenzae. Electrophoresis 21:411-429.
Lasonder E, Ishihama Y, Andersen JS, Vermunt AM, Pain A, Sauerwein RW, Eling WM, Hall N, Waters AP, Stunnenberg HG, Mann M. 2002. Analysis of the Plasmodium falciparum proteome by high-accuracy mass spectrometry. Nature 419:537-542.
Lengqvist J, Uhlen K, Lehtio J. 2007. iTRAQ compatibility of peptide immobilized pH gradient isoelectric focusing. Proteomics 7:1746-1752.
Liang X, Bai J, Liu YH, Lubman DM. 1996. Characterization of SDS--PAGE-separated proteins by matrix-assisted laser desorption/ionization mass spectrometry. Anal Chem 68:1012-1018.
Link AJ, Eng J, Schieltz DM, Carmack E, Mize GJ, Morris DR, Garvik BM, Yates JR, 3rd. 1999. Direct analysis of protein complexes using mass spectrometry. Nat Biotechnol 17:676-682.
Liu H, Sadygov R G, Yates III J R. 2004 A model for random sampling and estimation of relative protein abundance in shotgun proteomics. Anal. Chem., 76: 4193-4201.
Locke S, Figeys D. 2000. Techniques for the optimization of proteomic strategies based on head column stacking capillary electrophoresis. Anal Chem 72:2684-2689.
Loo RRO, Hayes R, Yang YN, Hung F, Ramachandran P, Kim N, Gunsalus R, Loo JA. 2005. Top-down, bottom-up, and side-to-side proteomics with virtual 2-D gels. International Journal of Mass Spectrometry 240:317-325.
Lopez MF, Patton WF, Utterback BL, Chung-Welch N, Barry P, Skea WM, Cambria RP. 1991. Effect of various detergents on protein migration in the second dimension of two-dimensional gels. Anal Biochem 199:35-44.
Lu X, Zhu H. 2005. Tube-gel digestion: a novel proteomic approach for high throughput analysis of membrane proteins. Mol Cell Proteomics 4:1948-1958.
Lu B, Motoyama A, Ruse C, Venable J, Yates III JR. 2008 Improving protein identification sensitivity by combining MS and MS/MS information for shotgun proteomics using LTQ-Orbitrap high mass accuracy data Anal.Chem., 80: 2018-2025.
Luche S, Lelong C, Diemer H, Van Dorsselaer A, Rabilloud T. 2007. Ultrafast coelectrophoretic fluorescent staining of proteins with carbocyanines. Proteomics 7: 3234-3244.
Lui M, Tempst P, Erdjument-Bromage H. 1996. Methodical analysis of protein-nitrocellulose interactions to design a refined digestion protocol. Anal Biochem 241:156-166.

Luo S, Wehr NB, Levine RL. 2006. Quantitation of protein on gels and blots by infrared fluorescence of Coomassie blue and Fast Green. Anal Biochem 350:233-238.
Luque-Garcia JL, Zhou G, Sun TT, Neubert TA. 2006. Use of nitrocellulose membranes for protein characterization by matrix-assisted laser desorption/ionization mass spectrometry. Anal Chem 78:5102-5108.
Lyubimova T, Caglio S, Gelfi C, Righetti PG, Rabilloud T. 1993. Photopolymerization of Polyacrylamide Gels with Methylene-Blue. Electrophoresis 14:40-50.
Macfarlane DE. 1983. Use of benzyldimethyl-n-hexadecylammonium chloride ("16-BAC"), a cationic detergent, in an acidic polyacrylamide gel electrophoresis system to detect base labile protein methylation in intact cells. Anal Biochem 132:231-235.
Macfarlane DE. 1989. Two dimensional benzyldimethyl-n-hexadecylammonium chloride----sodium dodecyl sulfate preparative polyacrylamide gel electrophoresis: a high capacity high resolution technique for the purification of proteins from complex mixtures. Anal Biochem 176:457-463.
MacGillivray AJ, Rickwood D. 1974. The heterogeneity of mouse-chromatin nonhistone proteins as evidenced by two-dimensional polyacrylamide-gel electrophoresis and ion-exchange chromatography. Eur J Biochem 41:181-190.
Mackintosh JA, Choi HY, Bae SH, Veal DA, Bell PJ, Ferrari BC, Van Dyk DD, Verrills NM, Paik YK, Karuso P. 2003. A fluorescent natural product for ultra sensitive detection of proteins in one-dimensional and two-dimensional gel electrophoresis. Proteomics 3:2273-2288.
Malmstrom J, Lee H, Nesvizhskii AI, Shteynberg D, Mohanty S, Brunner E, Ye M, Weber G, Eckerskorn C, Aebersold R. 2006. Optimized peptide separation and identification for mass spectrometry based proteomics via free-flow electrophoresis. J Proteome Res 5:2241-2249.
Malone JP, Radabaugh MR, Leimgruber RM, Gerstenecker GS. 2001. Practical aspects of fluorescent staining for proteomic applications. Electrophoresis 22:919-932.
Matsudaira P. 1987. Sequence from picomole quantities of proteins electroblotted onto polyvinylidene difluoride membranes. J Biol Chem 262:10035-10038.
Mazzeo JR, Martineau JA, Krull IS. 1993. Peptide mapping using EOF-driven capillary isoelectric focusing. Anal Biochem 208:323-329.
Mets LJ, Bogorad L. 1974. Two-dimensional polyacrylamide gel electrophoresis: an improved method for ribosomal proteins. Anal Biochem 57:200-210.
Moritz RL, Ji H, Schutz F, Connolly LM, Kapp EA, Speed TP, Simpson RJ. 2004. A proteome strategy for fractionating proteins and peptides using continuous free-flow electrophoresis coupled off-line to reversed-phase high-performance liquid chromatography. Anal Chem 76:4811-4824.
Nakamura K, Okuya Y, Katahira M, Yoshida S, Wada S, Okuno M. 1992. Analysis of tubulin isoforms by two-dimensional gel electrophoresis using SDS-polyacrylamide gel electrophoresis in the first dimension. J Biochem Biophys Methods 24:195-203.
Navarre C, Degand H, Bennett KL, Crawford JS, Mortz E, Boutry M. 2002. Subproteomics: identification of plasma membrane proteins from the yeast Saccharomyces cerevisiae. Proteomics 2:1706-1714.
Neuhoff V, Arold N, Taube D, Ehrhardt W. 1988. Improved staining of proteins in polyacrylamide gels including isoelectric focusing gels with clear background at nanogram sensitivity using Coomassie Brilliant Blue G-250 and R-250. Electrophoresis 9:255-262.
Nielsen ML, Savitski MM, Zubarev RA. 2006 Extent of modifications in human proteome samples and their effect on dynamic range of analysis in shotgun proteomics
Mol. Cell. Proteomics 5: 2384-2391
Novakova Z, Man P, Novak P, Hozak P, Hodny Z. 2006. Separation of nuclear protein complexes by blue native polyacrylamide gel electrophoresis. Electrophoresis 27:1277-1287.
O'Farrell PH. 1975. High resolution two-dimensional electrophoresis of proteins. J Biol Chem 250:4007-4021.
O'Farrell PZ, Goodman HM, O'Farrell PH. 1977. High resolution two-dimensional electrophoresis of basic as well as acidic proteins. Cell 12:1133-1141.
Ogorzalek Loo RR, Mitchell C, Stevenson TI, Martin SA, Hines WM, Juhasz P, Patterson DH, Peltier JM, Loo JA, Andrews PC. 1997. Sensitivity and mass accuracy for proteins analyzed directly from polyacrylamide gels: implications for proteome mapping. Electrophoresis 18:382-390.
Orrick LR, Olson MO, Busch H. 1973. Comparison of nucleolar proteins of normal rat liver and Novikoff hepatoma ascites cells by two-dimensional polyacrylamide gel electrophoresis. Proc Natl Acad Sci U S A 70:1316-1320.
Ortiz ML, Calero M, Fernandez Patron C, Patron CF, Castellanos L, Mendez E. 1992. Imidazole-SDS-Zn reverse staining of proteins in gels containing or not SDS and microsequence of individual unmodified electroblotted proteins. FEBS Lett 296:300-304.
Oses-Prieto JA, Zhang X, Burlingame AL. 2007. Formation of epsilon-formyllysine on silver-stained proteins: implications for assignment of isobaric dimethylation sites by tandem mass spectrometry. Mol Cell Proteomics 6:181-192.
Pappin DJ, Hojrup P, Bleasby AJ. 1993. Rapid identification of proteins by peptide-mass fingerprinting. Curr Biol 3:327-332.
Patton WF, Chung-Welch N, Lopez MF, Cambria RP, Utterback BL, Skea WM. 1991. Tris-tricine and Tris-borate buffer systems provide better estimates of human mesothelial cell intermediate filament protein molecular weights than the standard Tris-glycine system. Anal Biochem 197:25-33.
Peng J, Elias JE, Thoreen CC, Licklider LJ, Gygi SP. 2003. Evaluation of multidimensional chromatography coupled with tandem mass spectrometry (LC/LC-MS/MS) for large-scale protein analysis: the yeast proteome. J Proteome Res 2:43-50.
Perkins DN, Pappin DJ, Creasy DM, Cottrell JS. 1999. Probability-based protein identification by searching sequence databases using mass spectrometry data. Electrophoresis 20:3551-3567.
Petritis K, Kangas LJ, Ferguson PL, Anderson GA, Pasa-Tolic L, Lipton MS, Auberry KJ, Strittmatter EF, Shen Y, Zhao R, Smith RD. 2003. Use of artificial neural networks for the accurate prediction of peptide liquid chromatography elution times in proteome analyses. Anal Chem 75:1039-1048.
Phelps DS. 1984. Electrophoretic transfer of proteins from fixed and stained gels. Anal Biochem 141:409-412.
Rabilloud T. 1990. Mechanisms of protein silver staining in polyacrylamide gels: a 10-year synthesis. Electrophoresis 11:785-794.
Rabilloud T. 1998. Use of thiourea to increase the solubility of membrane proteins in two-dimensional electrophoresis. Electrophoresis 19:758-760.
Rabilloud T, Heller M, Gasnier F, Luche S, Rey C, Aebersold R, Benahmed M, Louisot P, Lunardi J. 2002. Proteomics analysis of cellular response to oxidative stress - Evidence for in vivo overoxidation of peroxiredoxins at their active site. Journal of Biological Chemistry 277:19396-19401.
Rabilloud T, Strub JM, Luche S, van Dorsselaer A, Lunardi J. 2001. Comparison between Sypro Ruby and ruthenium II tris (bathophenanthroline disulfonate) as fluorescent stains for protein detection in gels. Proteomics 1:699-704.
Radola BJ. 1973. Isoelectric focusing in layers of granulated gels. I. Thin-layer isoelectric focusing of proteins. Biochim Biophys Acta 295:412-428.
Radola BJ. 1975. Isoelectric focusing in layers of granulated gels. II. Preparative isoelectric focusing. Biochim Biophys Acta 386:181-195.
Rais I, Karas M, Schagger H. 2004. Two-dimensional electrophoresis for the isolation of integral membrane proteins and mass spectrometric identification. Proteomics 4:2567-2571.
Reid GE, Ji H, Eddes JS, Moritz RL, Simpson RJ. 1995. Nonreducing two-dimensional polyacrylamide gel electrophoretic analysis of human colonic proteins. Electrophoresis 16:1120-1130.
Reynolds JA, Tanford C. 1970. Binding of dodecyl sulfate to proteins at high binding ratios. Possible implications for the state of proteins in biological membranes. Proc Natl Acad Sci U S A 66:1002-1007.
Richert S, Luche S, Chevallet M, Van Dorsselaer A, Leize-Wagner E, Rabilloud T. 2004. About the mechanism of interference of silver staining with peptide mass spectrometry. Proteomics 4:909-916.
Rigaut G, Shevchenko A, Rutz B, Wilm M, Mann M, Seraphin B. 1999. A generic protein purification method for protein complex characterization and proteome exploration. Nat Biotechnol 17:1030-1032.
Righetti PG. 1989. Of Matrices and Men. Journal of Biochemical and Biophysical Methods 19:1-20.
Righetti PG, Barzaghi B, Luzzana M, Manfredi G, Faupel M. 1987. A Horizontal Apparatus for Isoelectric Protein-Purification in a Segmented Immobilized Ph Gradient. Journal of Biochemical and Biophysical Methods 15:189-198.
Righetti PG, Caglio S. 1993. On the kinetics of monomer incorporation into polyacrylamide gels, as investigated by capillary zone electrophoresis. Electrophoresis 14:573-582.
Righetti PG, Castagna A, Herbert B, Candiano G. 2005. How to bring the "unseen" proteome to the limelight via electrophoretic pre-fractionation techniques. Biosci Rep 25:3-17.
Righetti PG, Gianazza E. 1987. Isoelectric focusing in immobilized pH gradients: theory and newer methodology. Methods Biochem Anal 32:215-278.
Ros A, Faupel M, Mees H, Oostrum J, Ferrigno R, Reymond F, Michel P, Rossier JS, Girault HH. 2002. Protein purification by Off-Gel electrophoresis. Proteomics 2:151-156.
Rosenfeld J, Capdevielle J, Guillemot JC, Ferrara P. 1992. In-gel digestion of proteins for internal sequence analysis after one- or two-dimensional gel electrophoresis. Anal Biochem 203:173-179.
Sanchez JC, Ravier F, Pasquali C, Frutiger S, Paquet N, Bjellqvist B, Hochstrasser DF, Hughes GJ. 1992. Improving the Detection of Proteins after Transfer to Polyvinylidene Difluoride Membranes. Electrophoresis 13:715-717.
Santoni V, Molloy M, Rabilloud T. 2000. Membrane proteins and proteomics: Un amour impossible? Electrophoresis 21:1054-1070.
Schagger H, von Jagow G. 1987. Tricine-sodium dodecyl sulfate-polyacrylamide gel electrophoresis for the separation of proteins in the range from 1 to 100 kDa. Anal Biochem 166:368-379.
Schagger H, von Jagow G. 1991. Blue native electrophoresis for isolation of membrane protein complexes in enzymatically active form. Anal Biochem 199:223-231.
Scherl A, Coute Y, Deon C, Calle A, Kindbeiter K, Sanchez JC, Greco A, Hochstrasser D, Diaz JJ. 2002. Functional proteomic analysis of human nucleolus. Molecular Biology of the Cell 13:4100-4109.
Scherl A, Francois P, Charbonnier Y, Deshusses JM, Koessler T, Huyghe A, Bento M, Stahl-Zeng J, Fischer A, Masselot A, Vaezzadeh A, Galle F, Renzoni A, Vaudaux P, Lew D, Zimmermann-Ivol CG, Binz PA, Sanchez JC, Hochstrasser DF, Schrenzel J. 2006. Exploring glycopeptide-resistance in Staphylococcus aureus: a combined proteomics and transcriptomics approach for the identification of resistance-related markers. BMC Genomics 7:296.
Schirle M, Heurtier MA, Kuster B. 2003. Profiling core proteomes of human cell lines by one-dimensional PAGE and liquid chromatography-tandem mass spectrometry. Molecular & Cellular Proteomics 2:1297-1305.
Schluesener D, Rogner M, Poetsch A. 2007. Evaluation of two proteomics technologies used to screen the membrane proteomes of wild-type Corynebacterium glutamicum and an L-lysine-producing strain. Anal Bioanal Chem 389:1055-1064.
Schmitt LM. 2001. Theory of genetic algorithms. Theoretical Computer Science 259:1-61.
Sevinsky JR, Brown KJ, Cargile BJ, Bundy JL, Stephenson JL, Jr. 2007. Minimizing back exchange in 18O/16O quantitative proteomics experiments by incorporation of immobilized trypsin into the initial digestion step. Anal Chem 79:2158-2162.
Sevinsky JR, Cargile BJ, Bunger MK, Meng F, Yates NA, Hendrickson RC, Stephenson JL, Jr. 2008. Whole genome searching with shotgun proteomic data: applications for genome annotation. J

Proteome Res 7:80-88.
Shang TQ, Ginter JM, Johnston MV, Larsen BS, McEwen CN. 2003. Carrier ampholyte-free solution isoelectric focusing as a prefractionation method for the proteomic analysis of complex protein mixtures. Electrophoresis 24:2359-2368.
Shapiro AL, Vinuela E, Maizel JV, Jr. 1967. Molecular weight estimation of polypeptide chains by electrophoresis in SDS-polyacrylamide gels. Biochem Biophys Res Commun 28:815-820.
Shen Y, Berger SJ, Anderson GA, Smith RD. 2000. High-efficiency capillary isoelectric focusing of peptides. Anal Chem 72:2154-2159.
Shimura K, Kamiya K, Matsumoto H, Kasai K. 2002. Fluorescence-labeled peptide pI markers for capillary isoelectric focusing. Anal Chem 74:1046-1053.
Shimura K, Zhi W, Matsumoto H, Kasai K. 2000. Accuracy in the determination of isoelectric points of some proteins and a peptide by capillary isoelectric focusing: utility of synthetic peptides as isoelectric point markers. Anal Chem 72:4747-4757.
Siemankowski RF, Giambalvo A, Dreizen P. 1978. A new analytical procedure for two-dimensional electrophoresis of cellular proteins: comparison of protein compositions of parent strain and a K+-accumulation mutant of E. coli. Physiol Chem Phys 10:415-434.
SimoAlfonso E, Gelfi C, Sebastiano R, Citterio A, Righetti PG. 1996. Novel acrylamido monomers with higher hydrophilicity and improved hydrolytic stability .2. Properties of N-acryloylaminopropanol. Electrophoresis 17:732-737.
Simpson DC, Smith RD. 2005. Combining capillary electrophoresis with mass spectrometry for applications in proteomics. Electrophoresis 26:1291-1305.
Steinberg TH, Haugland RP, Singer VL. 1996. Applications of SYPRO Orange and SYPRO Red protein gel stains. Analytical Biochemistry 239:238-245.
Steinberg TH, Lauber WM, Berggren K, Kemper C, Yue S, Patton WF. 2000. Fluorescence detection of proteins in sodium dodecyl sulfate-polyacrylamide gels using environmentally benign, nonfixative, saline solution. Electrophoresis 21:497-508.
Switzer RC, 3rd, Merril CR, Shifrin S. 1979. A highly sensitive silver stain for detecting proteins and peptides in polyacrylamide gels. Anal Biochem 98:231-237.
Szponarski W, Delom F, Sommerer N, Rossignol M, Gibrat R. 2007. Separation, identification, and profiling of membrane proteins by GFC/IEC/SDS-PAGE and MALDI TOF MS. Methods Mol Biol 355:267-278.
Szponarski W, Sommerer N, Boyer JC, Rossignol M, Gibrat R. 2004. Large-scale characterization of integral proteins from Arabidopsis vacuolar membrane by two-dimensional liquid chromatography. Proteomics 4:397-406.
Tabb DL, McDonald WH, Yates JR, 3rd. 2002. DTASelect and Contrast: tools for assembling and comparing protein identifications from shotgun proteomics. J Proteome Res 1:21-26.
Tan A, Pashkova A, Zang L, Foret F, Karger BL. 2002. A miniaturized multichamber solution isoelectric focusing device for separation of protein digests. Electrophoresis 23:3599-3607.
Tarroux P, Vincens P, Rabilloud T. 1987. Hermes - a 2nd Generation Approach to the Automatic-Analysis of Two-Dimensional Electrophoresis Gels .5. Data-Analysis. Electrophoresis 8:187-199.
Tastet C, Lescuyer P, Diemer H, Luche S, van Dorsselaer A, Rabilloud T. 2003. A versatile electrophoresis system for the analysis of high- and low-molecular-weight proteins. Electrophoresis 24:1787-1794.
Towbin H, Staehelin T, Gordon J. 1979. Electrophoretic transfer of proteins from polyacrylamide gels to nitrocellulose sheets: procedure and some applications. Proc Natl Acad Sci U S A 76:4350-4354.
Tsugita A, Miyazaki K, Nabetani T, Nozawa T, Kamo M, Kawakami T. 2001. Application of chemical selective cleavage methods to analyze post-translational modification in proteins. Proteomics 1:1082-1091.
Tuszynski GP, Buck CA, Warren L. 1979. A two-dimensional polyacrylamide gel electrophoresis

(PAGE) system using sodium dodecyl sulfate-PAGE in the first dimension. Anal Biochem 93:329-338.
Unlu M, Morgan ME, Minden JS. 1997. Difference gel electrophoresis: a single gel method for detecting changes in protein extracts. Electrophoresis 18:2071-2077.
van Oostveen I, Ducret A, Aebersold R. 1997. Colloidal silver staining of electroblotted proteins for high sensitivity peptide mapping by liquid chromatography-electrospray ionization tandem mass spectrometry. Anal Biochem 247:310-318.
Vonguyen L, Wu J, Pawliszyn J. 1994. Peptide mapping of bovine and chicken cytochrome c by capillary isoelectric focusing with universal concentration gradient imaging. J Chromatogr B Biomed Appl 657:333-338.
Voris BP, Young DA. 1980. Very-high-resolution two-dimensional gel electrophoresis of proteins using giant gels. Anal Biochem 104:478-484.
Wang W, Guo T, Rudnick PA, Song T, Li J, Zhuang Z, Zheng W, Devoe DL, Lee CS, Balgley BM. 2007. Membrane proteome analysis of microdissected ovarian tumor tissues using capillary isoelectric focusing/reversed-phase liquid chromatography-tandem MS. Anal Chem 79:1002-1009.
Wang Y, Hancock WS, Weber G, Eckerskorn C, Palmer-Toy D. 2004. Free flow electrophoresis coupled with liquid chromatography-mass spectrometry for a proteomic study of the human cell line (K562/CR3). J Chromatogr A 1053:269-278.
Washburn MP, Wolters D, Yates JR. 2001. Large-scale analysis of the yeast proteome by multidimensional protein identification technology. Nature Biotechnology 19:242-247.
Weber K, Kuter DJ. 1971. Reversible denaturation of enzymes by sodium dodecyl sulfate. J Biol Chem 246:4504-4509.
Wen L, Tweten RK, Isackson PJ, Iandolo JJ, Reeck GR. 1983. Ionic interactions between proteins in nonequilibrium pH gradient electrophoresis: histones affect the migration of high mobility group nonhistone chromatin proteins. Anal Biochem 132:294-304.
West MHP, Wu RS, Bonner WM. 1984. Polyacrylamide-Gel Electrophoresis of Small Peptides. Electrophoresis 5:133-138.
White IR, Pickford R, Wood J, Skehel JM, Gangadharan B, Cutler P. 2004. A statistical comparison of silver and SYPRO Ruby staining for proteomic analysis. Electrophoresis 25:3048-3054.
Wilkins MR, Gasteiger E, Sanchez JC, Bairoch A, Hochstrasser DF. 1998. Two-dimensional gel electrophoresis for proteome projects: The effects of protein hydrophobicity and copy number. Electrophoresis 19:1501-1505.
Williams TI, Combs JC, Thakur AP, Strobel HJ, Lynn BC. 2006. A novel Bicine running buffer system for doubled sodium dodecyl sulfate - polyacrylamide gel electrophoresis of membrane proteins. Electrophoresis 27:2984-2995.
Wiltfang J, Arold N, Neuhoff V. 1991. A new multiphasic buffer system for sodium dodecyl sulfate-polyacrylamide gel electrophoresis of proteins and peptides with molecular masses 100,000-1000, and their detection with picomolar sensitivity. Electrophoresis 12:352-366.
Winkler C, Denker K, Wortelkamp S, Sickmann A. 2007. Silver- and Coomassie-staining protocols: detection limits and compatibility with ESI MS. Electrophoresis 28:2095-2099.
Wise GE, Lin F. 1991. Transfer of silver-stained proteins from polyacrylamide gels to polyvinylidene difluoride membranes. J Biochem Biophys Methods 22:223-231.
Witze ES, Old WM, Resing KA, Ahn NG. 2007. Mapping protein post-translational modifications with mass spectrometry. Nature Methods 4:798-806.
Xiao Z, Conrads TP, Lucas DA, Janini GM, Schaefer CF, Buetow KH, Issaq HJ, Veenstra TD. 2004. Direct ampholyte-free liquid-phase isoelectric peptide focusing: application to the human serum proteome. Electrophoresis 25:128-133.
Xie H, Bandhakavi S, Griffin TJ. 2005. Evaluating preparative isoelectric focusing of complex peptide mixtures for tandem mass spectrometry-based proteomics: a case study in profiling chromatin-enriched subcellular fractions in Saccharomyces cerevisiae. Anal Chem 77:3198-3207.

Xie H, Bandhakavi S, Roe MR, Griffin TJ. 2007. Preparative peptide isoelectric focusing as a tool for improving the identification of lysine-acetylated peptides from complex mixtures. J Proteome Res 6:2019-2026.

Yamaguchi Y, Miyagi, Y, Baba H. 2008ba Two-dimensional electrophoresis with cationic detergents: a powerful tool for the proteomic analysis of myelin proteins. Part 1: technical aspects of electrophoresis.J Neurosci Res. 86:755-765.

Yamaguchi Y, Miyagi, Y, Baba H. 2008b Two-dimensional electrophoresis with cationic detergents: a powerful tool for the proteomic analysis of myelin proteins. Part 2: analytical aspects J Neurosci Res. 86:766-775.

Yang Y, Zhang S, Howe K, Wilson DB, Moser F, Irwin D, Thannhauser TW. 2007. A comparison of nLC-ESI-MS/MS and nLC-MALDI-MS/MS for GeLC-based protein identification and iTRAQ-based shotgun quantitative proteomics. J Biomol Tech 18:226-237.

Yata M, Sato K, Ohtsuki K, Kawabata M. 1996. Fractionation of peptides in protease digests of proteins by preparative isoelectric focusing in the absence of added ampholyte: A biocompatible and low-cost approach referred to as autofocusing. Journal of Agricultural and Food Chemistry 44:76-79.

Yates JR, 3rd, Speicher S, Griffin PR, Hunkapiller T. 1993. Peptide mass maps: a highly informative approach to protein identification. Anal Biochem 214:397-408.

Young DA. 1984. Advantages of separations on "giant" two-dimensional gels for detection of physiologically relevant changes in the expression of protein gene-products. Clin Chem 30:2104-2108.

Yu W, Li Y, Deng C, Zhang X. 2006. Comprehensive two-dimensional separation in coupling of reversed-phase chromatography with capillary isoelectric focusing followed by MALDI-MS identification using on-target digestion for intact protein analysis. Electrophoresis 27:2100-2110.

Yuan X, Kuramitsu Y, Furumoto H, Zhang X, Hayashi E, Fujimoto M, Nakamura K. 2007. Nuclear protein profiling of Jurkat cells during heat stress-induced apoptosis by 2-DE and MS/MS. Electrophoresis 28:2018-2026.

Zewert T, Harrington M. 1992. Polyethyleneglycol methacrylate 200 as an electrophoresis matrix in hydroorganic solvents. Electrophoresis 13:824-831.

Zischka H, Weber G, Weber PJ, Posch A, Braun RJ, Buhringer D, Schneider U, Nissum M, Meitinger T, Ueffing M, Eckerskorn C. 2003. Improved proteome analysis of Saccharomyces cerevisiae mitochondria by free-flow electrophoresis. Proteomics 3:906-916.

Zuo X, Speicher DW. 2000. A method for global analysis of complex proteomes using sample prefractionation by solution isoelectrofocusing prior to two-dimensional electrophoresis. Anal Biochem 284:266-278.

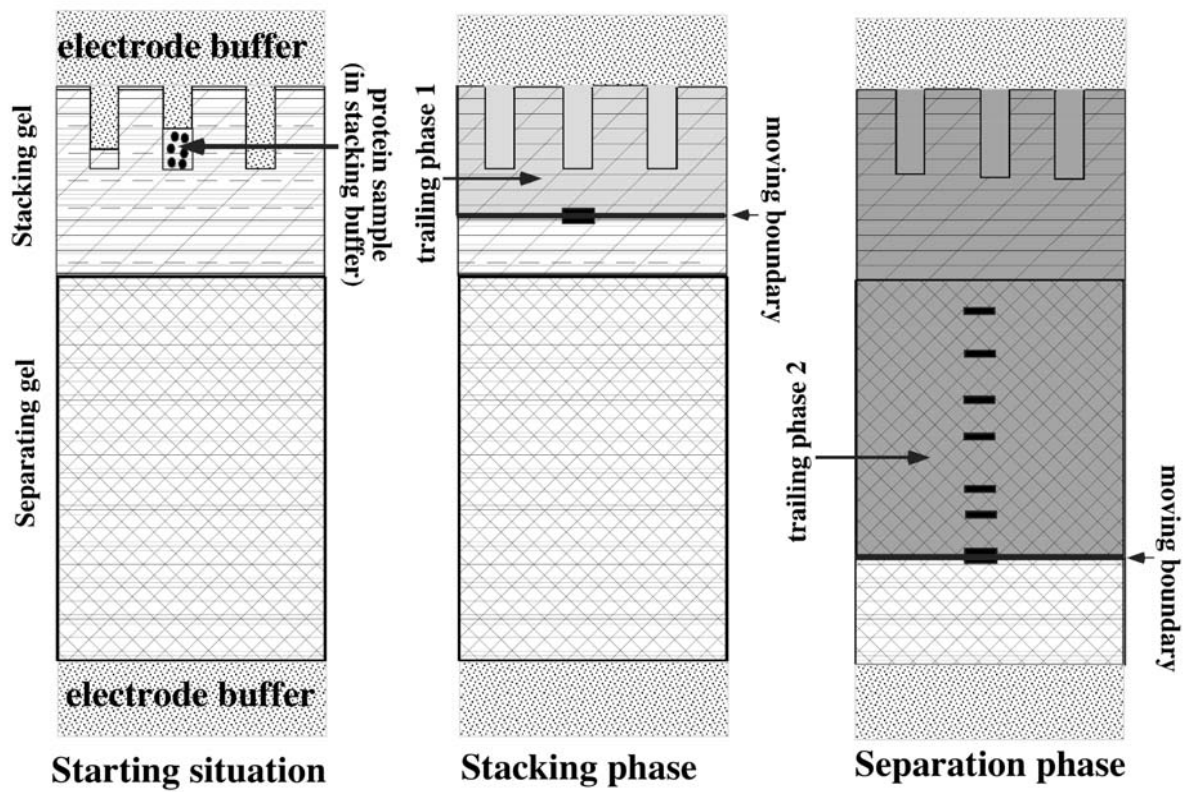

**Figure 1**:= Scheme of principle of discontinuous electrophoresis
A discontinuous gel (left panel) is made of three components

The electrode buffer contains the buffering ion (Tris in the case of the Laemmli system) and the trailing ion (Glycine in the Laemmli system)

The separating gel is a small pore size gel (tight hatching in the figure) cast in a buffer containing the buffering ion and the leading ion (resp. Tris-HCl buffer in the Laemmli system), The pH of this buffer is 8.8 in the Laemmli system.

The stacking gel is a large pore size gel (large hatching of the figure) cast in a buffer containing the buffering ion and the leading ion (resp. Tris-HCl buffer in the Laemmli system), The pH of this buffer is 6.8 in the Laemmli system.

The sample is loaded in the Tris-HCl pH 6.8 buffer

When the electrophoresis starts, (middle panel) the interaction of the electric field, electrode buffer and stacking gel buffer creates a moving boundary that separates the leading phase (i.e. the stacking gel buffer) from a trailing phase containing Tris, glycine, but whose pH is dictated by the interaction of the stacking and electrode buffer. In the case of the Laemmli buffer, the pH of this first trailing pahse is 8.3. The proteins are concentrated at the moving boundary (thick black band on the scheme).

When the electrophoresis proceeds in the separating gel, the trailing phase results from the interaction of the electrode buffer and separating gel buffer. Thus, the pH of the trailing phase changes (9.5 in the Laemmli system) and the speed of the moving boundary increases. The proteins are unstacked from the boundary and separated in sharp bands (black bands on the scheme). Any ion able to travel in the gel at the speed of the moving boundary stays in the stacked band (thick black band on the scheme).

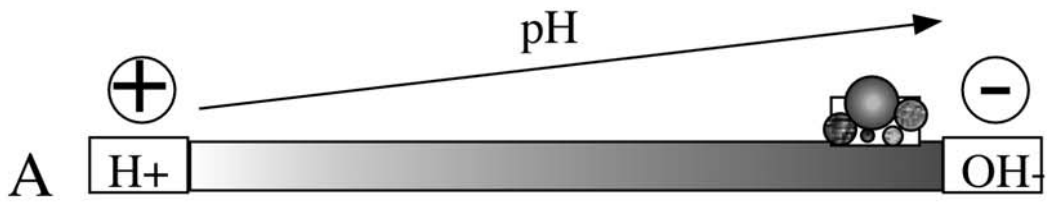

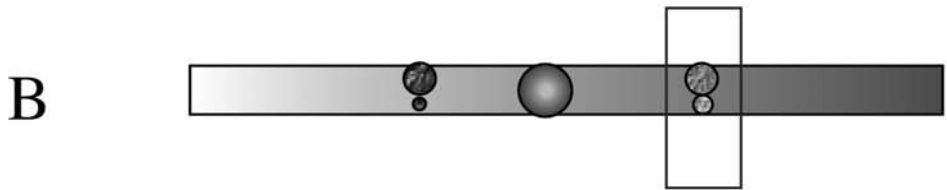

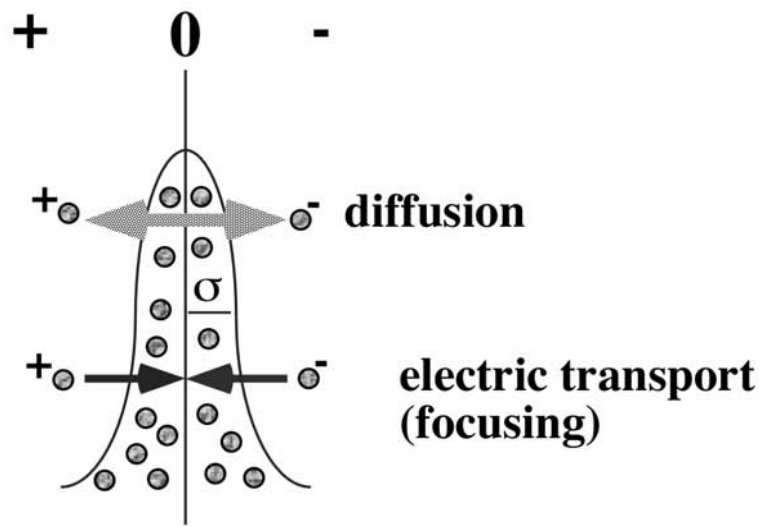

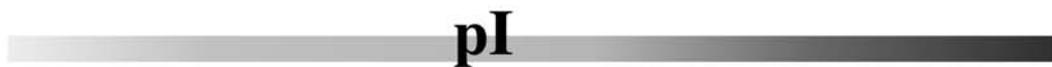

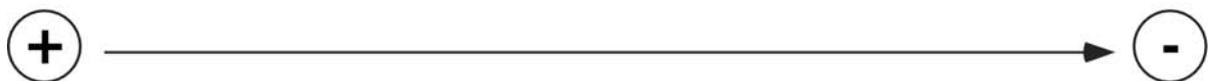

**Figure 2**: scheme of Isolectric Focusing (IEF)

A: starting situation: A polyacrylamide gel containing a pH gradient (either made with carrier ampholytes or immobilized pH gradient) is used as a support. The gel is connected to the electrodes (anode at the acidic side and cathode at the basic side) and the proteins are loaded on the gel

B: end situation: the proteins have been separated at their pI

C: detail of the focusing effect

The Gaussian peak represents the molecules focused at their pI. The diffusion tends to disperse molecules out of the pI position. If the molecules are brought to the acidic part of the gradient, they acquire a positive charge. The electric field tends to drive the molecules back to the pI. The converse (negative charges) is true on the basic side of the gradient. The electric field thus counteracts diffusion continuously, so that the width of the peak depends on the value of the electric field, and thin peaks are obtained

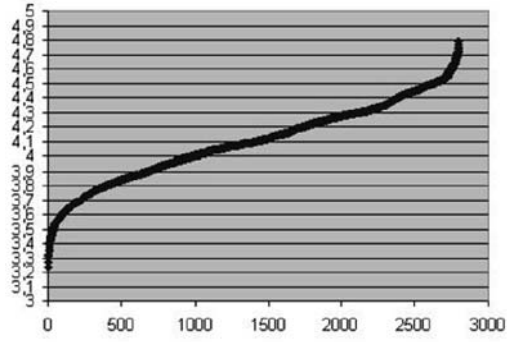
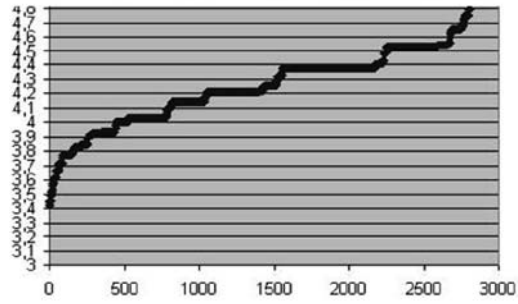

**Figure 3**: Theoretical peptides distribution of *S.aureus* in pI range 3.5-4.8 calculated by two different pI prediction algorithms.

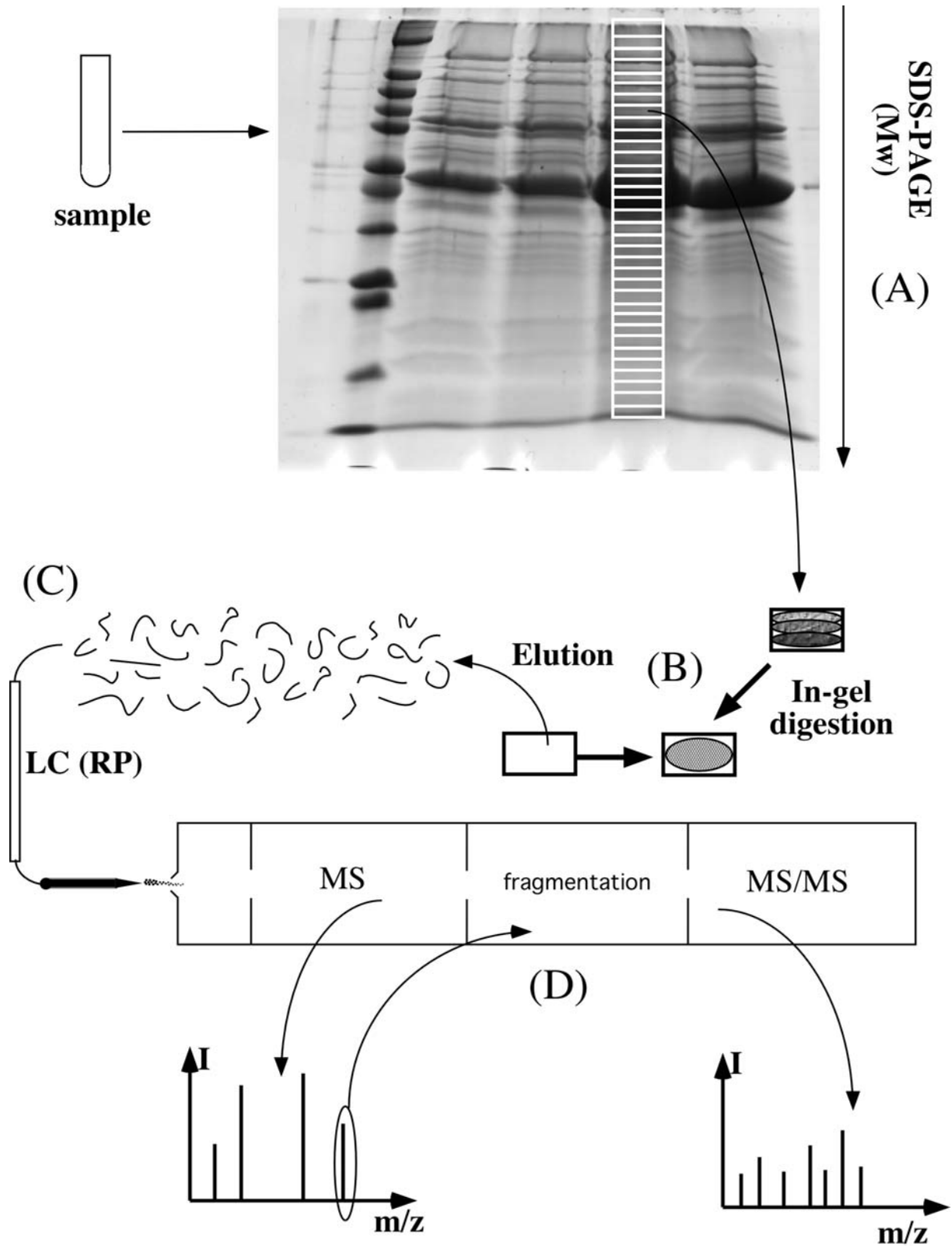

**Figure 4**: the GeLC method
The sample is separated first by SDS-PAGE (A). The lane of interest is divided into small gel pieces (usually without prior staining). Each gel piece, generally containing several proteins, is submitted to in-gel digestion (B). The peptides obtained from this digestion are elkuted, separated by RP-HPLC (C), and the peptides eluting from the column are identified by MS/MS (D).

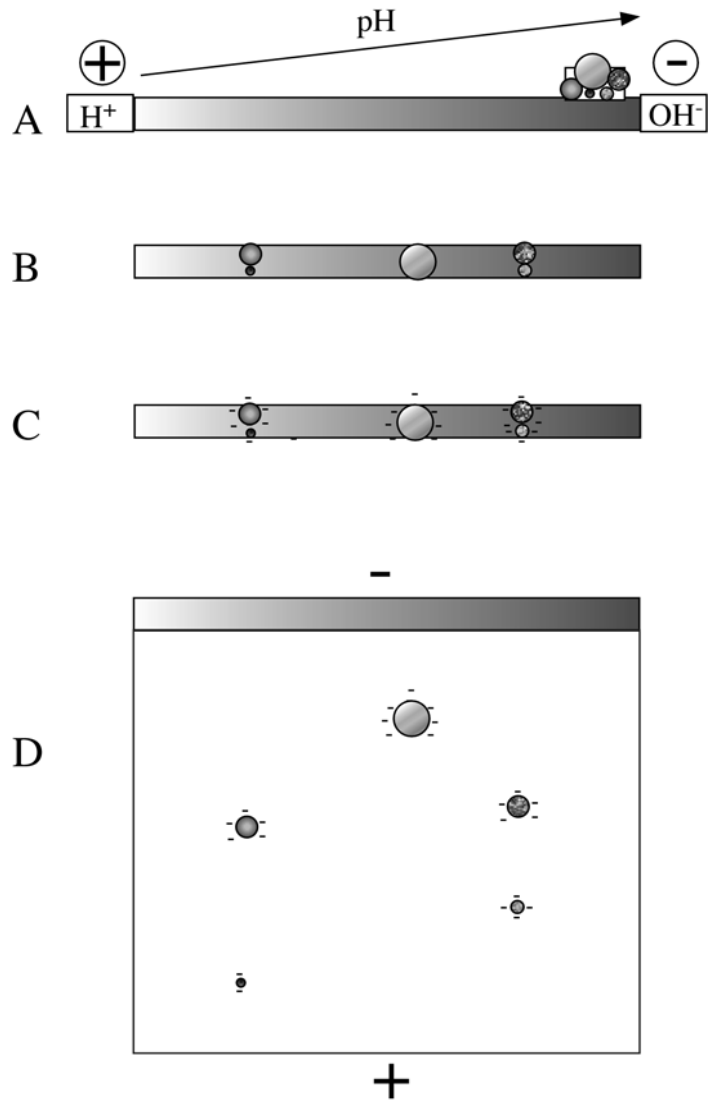

**Figure 5** : Scheme of principle of classical 2D electrophoresis (IEF/SDS-PAGE)
In the first step (A), the protein mixture is loaded on the IEF medium, i.e. a gel bearing a pH gradient. At the end of the IEF separation (B) the proteins have reach their respective isoelectric points, and it can be noted that different proteins may share the same pI. By definition, the proteins are electrically neutral at this position. The proteins are then saturated with SDS (C), which confers to them a strong negative charge. The SDS-saturated IEF gel is then loaded on top of a regular SDS-polyacrylamide gel, and the SDS separation is made (D).

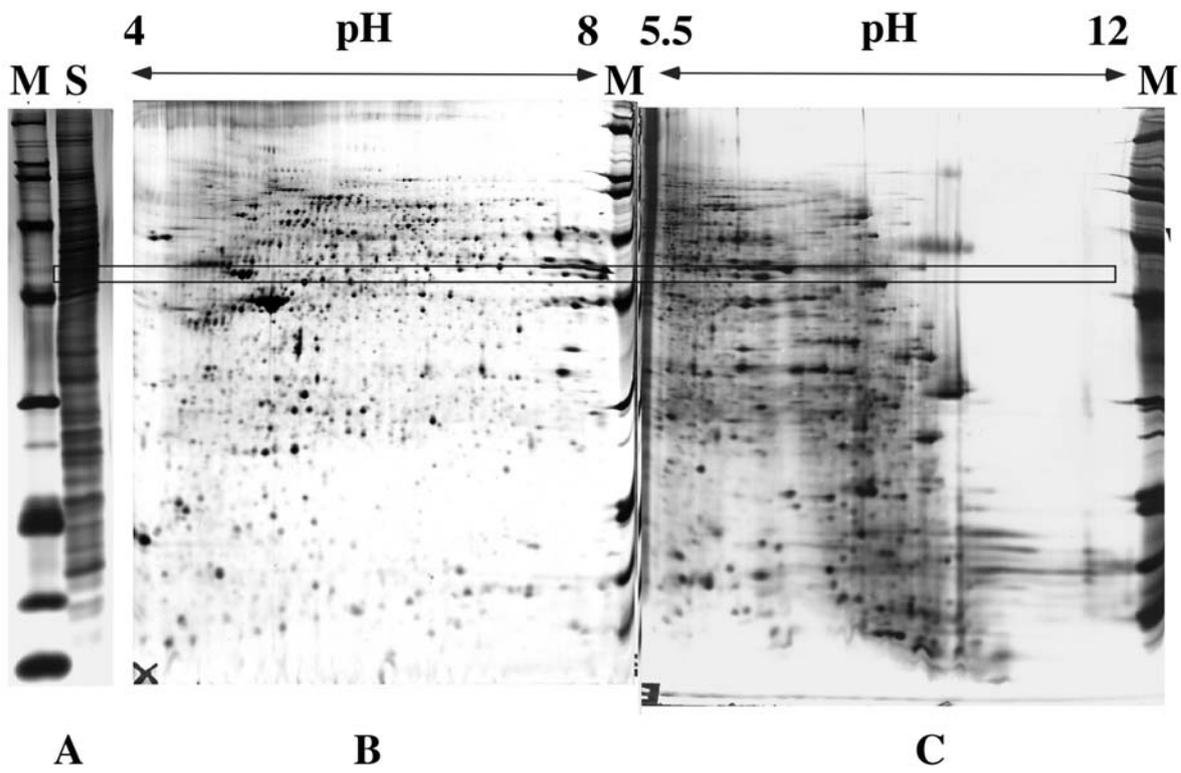

**Figure 6**: comparison of SDS gels with 2D gels
The sample (in this case human mitochondria) has been separated on a SDS gel (panel A) or on 2D gels with the indicated pH range (panels B and C) . M marks the position of the molecular weight markers, and S is the mitochondrial sample. The first dimensions of the 2D gels were IEF with immobilized pH gradients, and the SDS separation was in all cases carried out on a 11% acrylamide gel, using the Tris-taurine buffer system (see text)- The rectangle drawn on the gels allows one to appreciate how many different protein forms (i.e. how many protein spots) are present in a single band on the SDS gel.

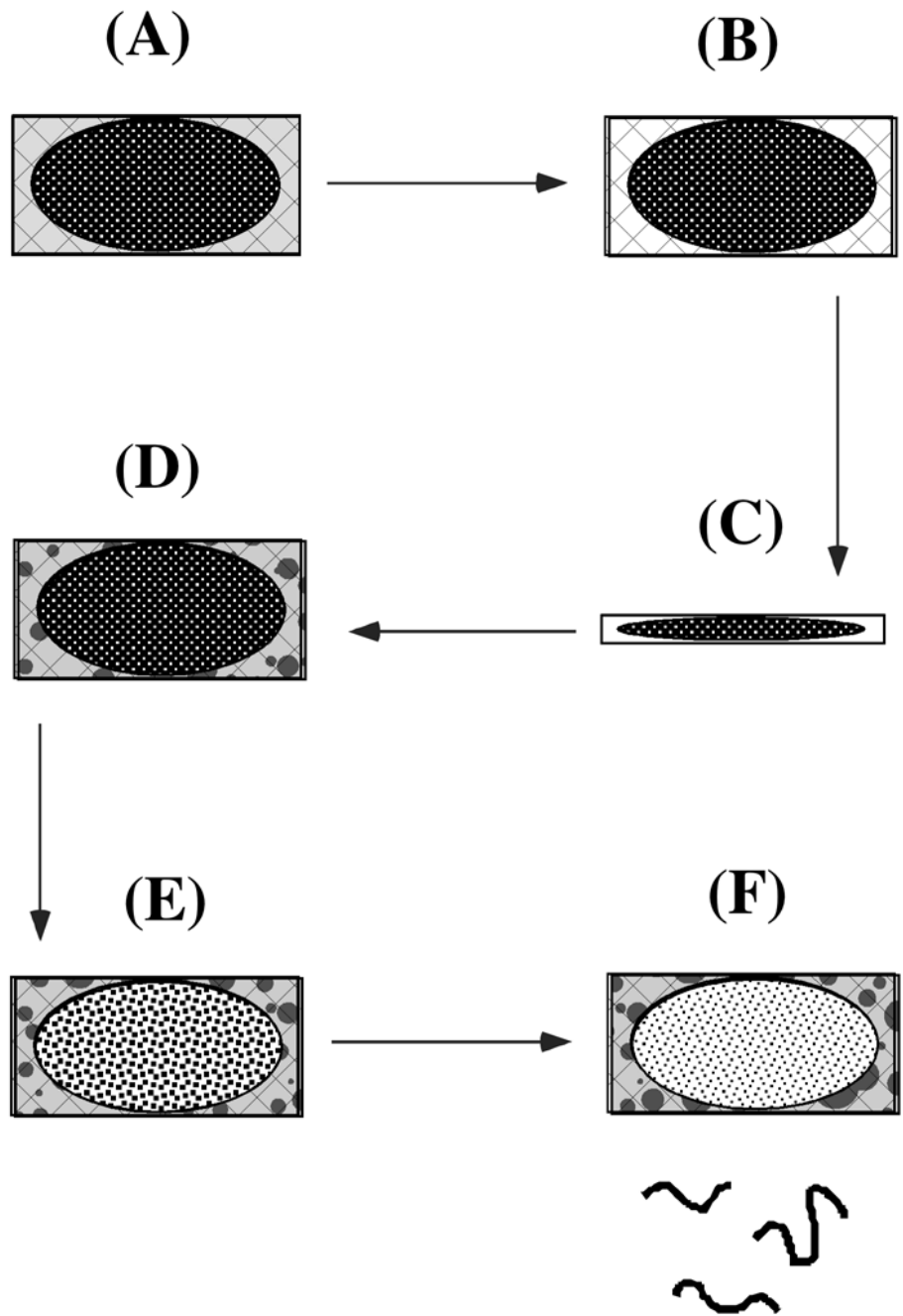

**Figure 7**: The in-gel digestion process
At the starting point (A), the protein to be digested (ellipsoid) is present in the gel (hatched network) containing the denaturing buffer components (symbolized by the grey color). The gels is the washed, so that these chemicals are eliminated (B, white gel background). The gel is the dehydrated by an organic solvent, usually acetonitrile, which induces gels shrinkage and protein precipitation (C). The gel is rehydrated in a protease solution, so that the protease (dark grey dots) enters in the gel network (D). The digestion proceeds, which degrades the protein, as symbolized by the change in the ellipsoid color (E). The peptides are eluted (F), but some material is probably not eluted and stays in the gel.

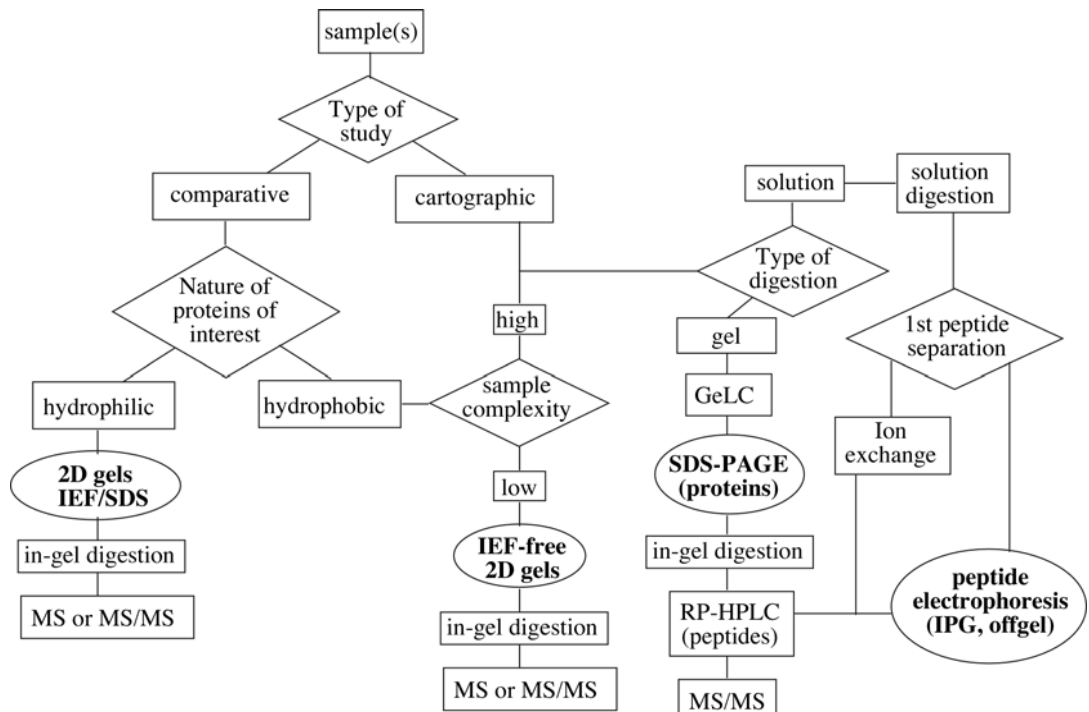

**Figure 8**: Tentative decision tree for proteomics studies
Starting from the sample(s), a series of questions (enclosed in diamond-shaped boxes) must be posed to devise the optimal proteomics strategy. Depending on the answers, various techniques can be used, and the electrophoretic techniques described in this review are shown in bold and in ellipsoid boxes. In the first question mark, the answer comparative/cartographic must be understood as the focus of the study. "Comparative" means that the focus will be put on the ability to compare several samples with a robust technique, while "cartographic" means that the focus will be put on the analysizs depth in terms of number of proteins identified. Other parameters, such as the average sequence coverage of the identified proteins at the mass spectrometry stage, or the importance of the post-translational modifications, have not been integrated in this decision tree to avoid over-complexification.